%% file: starcodex.tex
\documentclass[lettersize,journal]{IEEEtran}

\usepackage{amsmath,amsfonts,amssymb}
\usepackage{algorithmic}
\usepackage{algorithm}
\usepackage{array}

\usepackage[caption=false,font=normalsize,labelfont=sf,textfont=sf]{subfig}

\usepackage{textcomp}
\usepackage{stfloats}
\usepackage{url}
\usepackage{verbatim}
\usepackage{graphicx}
\usepackage{cite}
\usepackage{bm}
\usepackage{amsmath}
\usepackage[flushleft]{threeparttable}
\usepackage{booktabs}
\usepackage[T1]{fontenc} 
\usepackage{newtxmath}

\usepackage{color,soul}
\sethlcolor{yellow}

\begin{document}

\title{StarCodex: Dynamic Coding Harness for Starlink Measurement Analysis and Experiment Automation}
\author{Pengcheng Luo, Zhiming Shao, Bowen Zhang, Genke Yang, and Jian Chu\thanks{This work was supported by the National Major Science and Technology Project for Intelligent Manufacturing Systems and Robotics of China under Grant 2025ZD1602400.}
\thanks{Pengcheng Luo, Bowen Zhang, Genke Yang, and Jian Chu are with Ningbo Artificial Intelligence Institute, Shanghai Jiao Tong University, Ningbo 315000, China, and also with the School of Automation and Intelligent Sensing, Shanghai Jiao Tong University, Shanghai 200240, China, and the Key Laboratory of System Control and Information Processing, Ministry of Education of China, Shanghai 200240, China (e-mail: luopeng69131@sjtu.edu.cn, bwz96sco@sjtu.edu.cn, gkyang@sjtu.edu.cn, chujian@niii.com).}
\thanks{Zhiming Shao is with the School of Automation and Intelligent Sensing, Shanghai Jiao Tong University, Shanghai 200240, China (e-mail: zm.shao@sjtu.edu.cn).}
\thanks{The source code of StarCodex is available at: \protect\url{https://github.com/luopeng69131/StarCodex}.}
}

\maketitle

\begin{abstract}
Starlink and other low Earth orbit (LEO) satellite broadband systems are producing increasingly diverse measurement data across regions, time periods, and access conditions. These measurements are valuable for throughput prediction, adaptive bitrate (ABR) evaluation, and network experimentation, but converting continuously arriving data into reusable experimental evidence still relies heavily on manually developed analysis code and expert-guided data inspection and failure-case organization. This paper proposes \emph{StarCodex}, a dynamic coding harness for Starlink measurement analysis and experiment automation. StarCodex detects analysis gaps from the current measurement state, converts them into structured coding tasks, uses Codex to generate or repair executable analysis artifacts, and accepts artifacts through code, data-interface, measurement-semantics, and output validation. Experiments on real Starlink measurements show that StarCodex discovers 49 of 56 uncovered system-risk cases, attains higher average precision than the strongest predefined analysis baseline, and constructs a benchmark with denser and broader system-risk evidence. The generated prediction and replay artifacts further reveal prediction risks and quality-of-experience (QoE)--risk differences among ABR controllers. These results demonstrate the feasibility of using a dynamic coding harness to convert evolving Starlink measurements into validated analysis artifacts for automated experiment workflows.
\end{abstract}


\section{Introduction}
\subsection{Background and Motivation}
\IEEEPARstart{S}{tarlink} measurements are increasingly becoming the empirical basis for studying low Earth orbit (LEO) broadband systems~\cite{zhang2025eunomia,mohan2024multifaceted}. As deployments expand across regions, time periods, mobility states, and application scenarios, researchers are no longer faced with a single static trace collection, but with continuously arriving measurements that must be inspected, organized, and connected to downstream system experiments~\cite{li2025llm,michel2022first}. These measurements are therefore more than records of average throughput; they are operational evidence for judging whether a system can remain stable, which regions or periods expose risk, and which anomalies may affect throughput prediction, adaptive bitrate (ABR) control, and experimental evaluation~\cite{kassem2022browser,wu2025federated}.

\begin{figure}[!t]
  \centering
  \includegraphics[width=\linewidth]{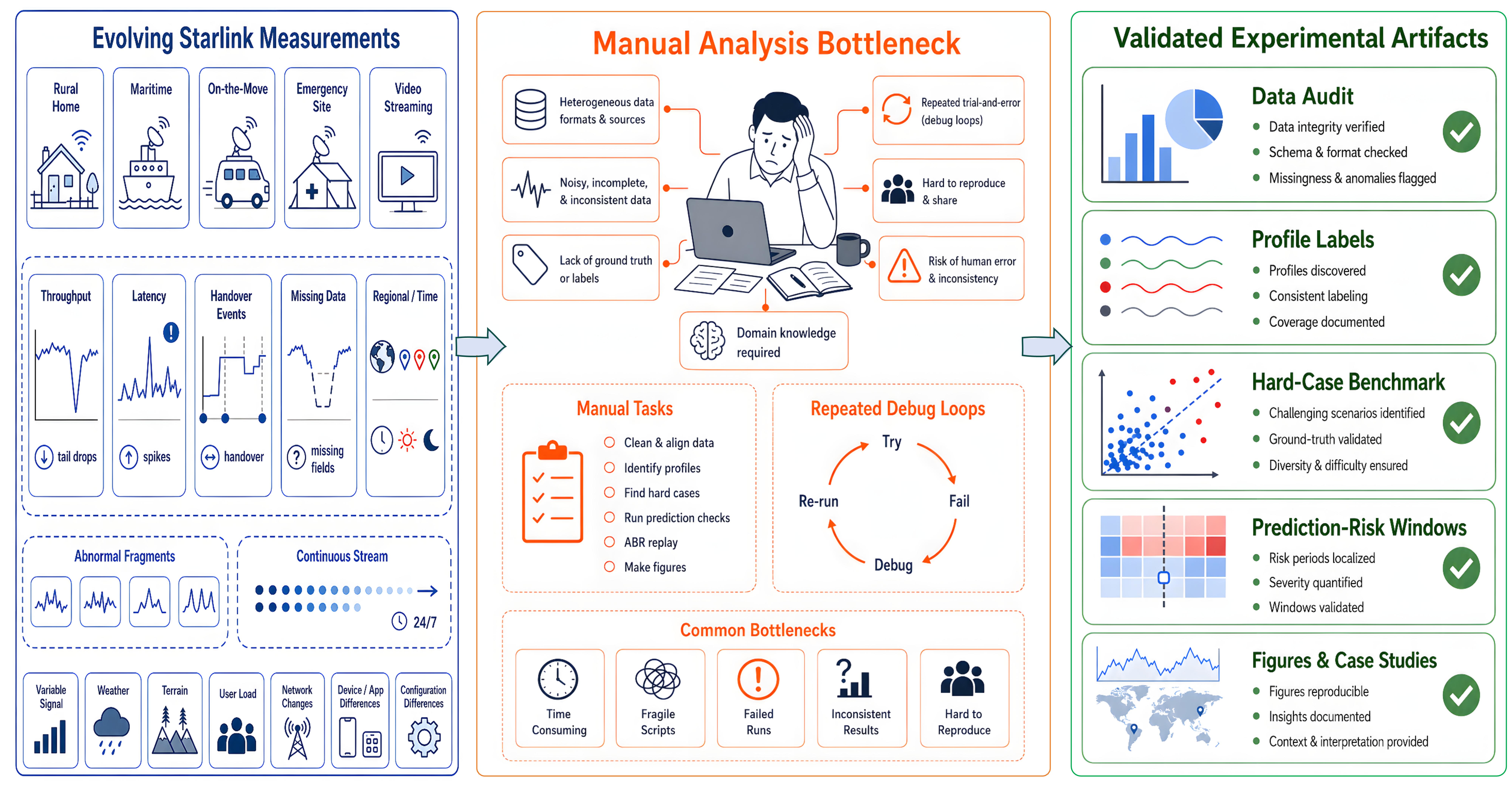}
  \caption{Starlink measurement-analysis problem. Evolving measurements expose heterogeneous access conditions, data-quality issues, and failure evidence; the desired output is a set of validated experimental artifacts that can support later system experiments.}
  \label{fig:intro-motivation}
\end{figure}

This measurement setting is continuously evolving rather than static and well aligned. New measurements may come from different regions, time periods, collection procedures, and link conditions; they may also include invalid intervals, missing fields, schema changes, or short-lived anomalies~\cite{mao2024intelligent,han2024cooperative}. Beyond these data-quality issues, many system-relevant risks are hidden by average statistics. A measurement segment with normal average throughput may contain a short capacity collapse; a throughput predictor with reasonable average error may overestimate capacity exactly in low-capacity periods; and an ABR controller that performs well on most sessions may still produce severe stalls on a small set of hard cases. The practical difficulty is therefore not only to run a fixed analysis script, but to continuously discover new data issues, risk patterns, failure cases, and experiment gaps as new Starlink measurements arrive.

Existing studies have made substantial progress in Starlink measurement, throughput prediction, and ABR control. Measurement studies characterize Starlink coverage, throughput, latency, and handover behavior~\cite{michel2022first,kassem2022browser,mohan2024multifaceted,liu2025starnet}. Prediction studies estimate future capacity or safe lower-bound capacity for streaming and network control~\cite{lv2022lumos,lv2024accurate,xie2026riskawaresafethroughputforecasting}. ABR studies improve playback quality through rule-based, optimization-based, or learned bitrate control~\cite{yin2015control,spiteri2020bola,mao2017neural,huang2019comyco,luo2025sabr,xie2026safesabr}. These works provide important system components and experimental foundations. However, the research workflow from newly collected Starlink measurements to reusable experimental data, risk features, hard-case benchmarks, figures, and case studies still requires substantial manual scripting, data inspection, debugging, and engineering judgment. As data volume increases and failure modes evolve, a central question emerges: how can changing Starlink measurements be converted into reliable analysis artifacts that support system experiments and conclusions? Fig.~\ref{fig:intro-motivation} illustrates this measurement-analysis bottleneck.

Large language models (LLMs) and coding harnesses such as Codex and Claude Code create a new opportunity for this problem~\cite{openai2026codex,anthropic2026claudecode}. Their strength is not limited to one-shot code generation. They can inspect existing scripts, respond to runtime errors, write new analysis programs, repair failed code, and revise experiments from feedback. This capability matches the workflow of Starlink measurement analysis: new data reveal new problems, new problems require new analysis code, and new analysis results may reshape the next experiment. Meanwhile, an executable script is not necessarily a meaningful analysis artifact, and a generated figure is not necessarily evidence of a real network risk. Therefore, Codex-like dynamic coding should be organized inside a process that checks code execution, data interfaces, measurement semantics, and output artifacts.

\begin{figure}[!t]
  \centering
  \includegraphics[width=\linewidth]{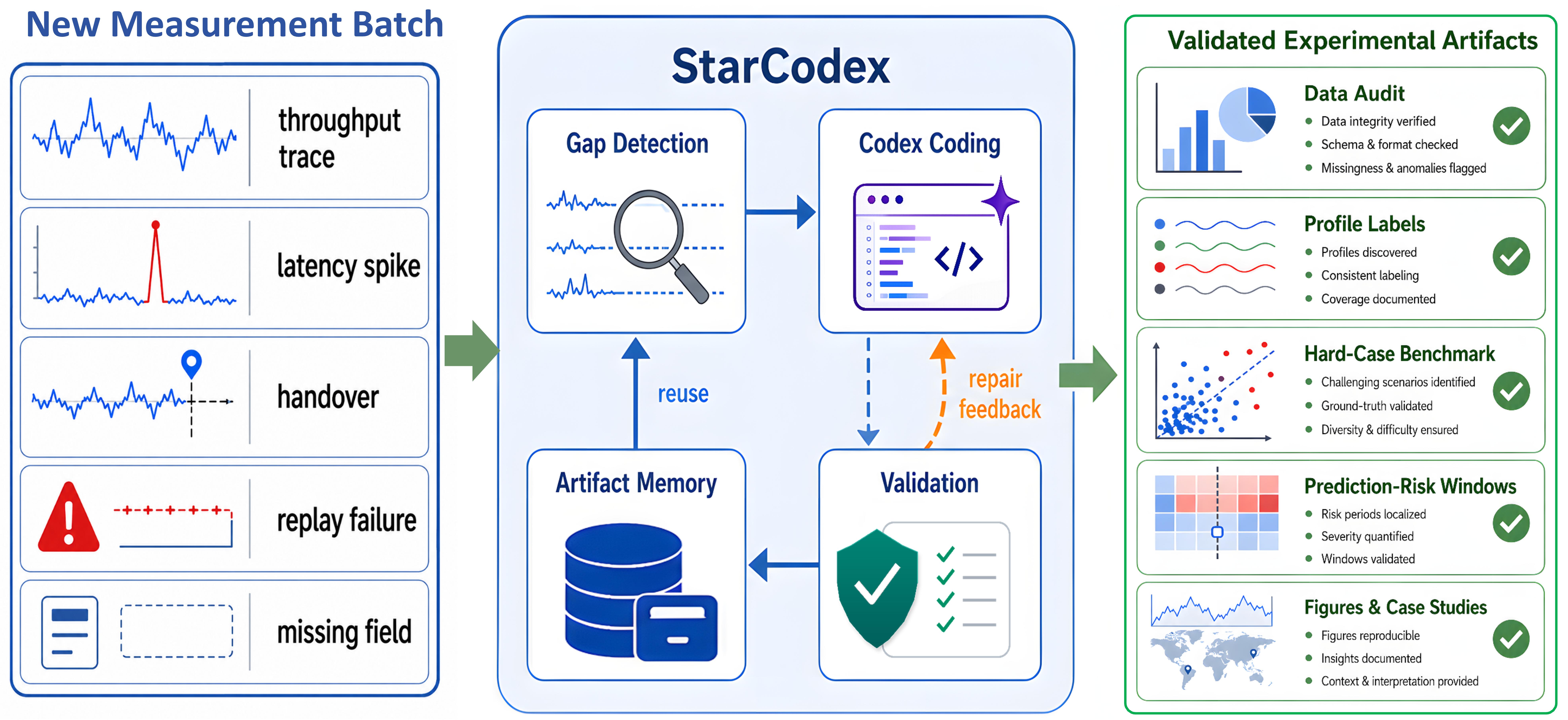}
  \caption{Conceptual overview of StarCodex.}
  \label{fig:starcodex-concept}
\end{figure}

Motivated by this observation, this paper proposes \emph{StarCodex}, a dynamic coding harness for Starlink measurement analysis and experiment automation. When new Starlink data arrive, StarCodex first uses the current analysis state to identify missing analyses, converts these gaps into structured coding tasks, invokes Codex to generate or repair the corresponding analysis programs, and accepts the resulting artifacts only after code-level, data-level, measurement-level, and output-level checks. Accepted artifacts are accumulated for later experiments, enabling reusable data auditing, profile discovery, feature extraction, failure diagnosis, benchmark construction, figure generation, and case-study analysis. Fig.~\ref{fig:starcodex-concept} gives a high-level view of this idea.

\subsection{Contributions}
The main contributions of this paper are summarized as follows:
\begin{itemize}

\item We propose \emph{StarCodex}, a dynamic coding harness for Starlink measurement analysis and experiment automation. StarCodex uses Codex to iteratively generate, test, and repair analysis artifacts, enabling the analysis workflow to adapt as new measurement risks, failure cases, and experiment needs appear.

\item We formulate Starlink measurement analysis as a \emph{measurement-to-artifact automation} workflow. It clarifies how continuously arriving Starlink measurements expose missing analyses and how these missing analyses can be turned into reusable evidence for experiment automation.

\item We evaluate the measurement-to-artifact workflow on real Starlink measurements using source-run-grouped cross-fitting. StarCodex discovers 49 of 56 uncovered system-risk cases, attains higher average precision than the strongest predefined analysis baseline, and constructs a benchmark with denser and broader system-risk evidence. These results demonstrate that dynamic coding can convert Starlink measurements into reusable evidence for system experiments.

\end{itemize}

\subsection{Organization}
The remainder of this paper is organized as follows. Section II reviews related work. Section III presents the measurement-to-artifact problem model. Section IV presents the StarCodex workflow. Section V evaluates StarCodex on real Starlink measurements. Section VI concludes the paper.

\section{Related Work}

This section reviews two research lines that motivate StarCodex. The first line studies Starlink measurement-driven system evaluation, where measurement traces become the basis for prediction, streaming, benchmark construction, and case analysis. The second line studies LLM-based coding harnesses and experiment automation, which provide the technical basis for dynamic coding with execution feedback.

\subsection{Starlink Measurement-Driven System Evaluation}

Starlink and other LEO satellite systems have attracted increasing attention as access networks for high-bandwidth and latency-sensitive applications. Measurement studies show that Starlink can provide broadband-like throughput while also exhibiting temporal variation, regional heterogeneity, access-path dynamics, handover-related behavior, and latency fluctuations~\cite{michel2022first,kassem2022browser,ma2023network,garcia2023multi,mohan2024multifaceted,liu2025starnet,izhikevich2024global}. These studies provide the empirical foundation for Starlink system research and, at the same time, show why raw measurements require careful interpretation: new runs may expose different throughput tails, latency patterns, data-quality issues, and failure cases.

Throughput prediction and ABR evaluation turn such measurements into downstream system evidence. Lumos, StarNet, and risk-aware safe-capacity forecasting estimate future capacity or conservative lower-bound capacity for streaming and network control~\cite{lv2022lumos,lv2024accurate,liu2025starnet,xie2026riskawaresafethroughputforecasting}. ABR and video-streaming studies evaluate buffer-based, optimization-based, and learned bitrate control under measured or replayed network conditions, including the Buffer Occupancy based Lyapunov Algorithm (BOLA), model predictive control (MPC)-style ABR, Pensieve, Comyco, SABR, SafeSABR, and Genet~\cite{yin2015control,spiteri2020bola,mao2017neural,huang2019comyco,luo2025sabr,xie2026safesabr,xia2022genet,zhao2023realtime,zhao2024dash}. These works show that prediction and playback conclusions depend not only on the controller or predictor being evaluated, but also on which measurement segments, risk windows, and replay cases are selected.

The preparation of these evaluation inputs is therefore part of the system experiment itself. A measurement segment must be audited, organized, labeled, linked to failure evidence, and compiled into an experimental artifact before it can support prediction-risk analysis, ABR replay, benchmark construction, or a case study. Existing measurement, prediction, and ABR studies provide important data sources and evaluation targets, but they usually start from prepared traces or manually constructed benchmarks. The remaining gap is how to continuously convert newly arriving Starlink measurements into validated artifacts that later experiments can reuse.

\subsection{LLM-Based Coding Harnesses for Experiment Automation}

LLMs have shown strong capabilities in reasoning, code generation, tool use, and feedback-driven improvement. ReAct and Toolformer connect language-model reasoning with external actions or tools, while Reflexion and OPRO use feedback to improve later attempts~\cite{yao2023react,schick2023toolformer,shinn2023reflexion,yang2024opro}. Software-engineering benchmarks and agents, including SWE-bench, SWE-agent, AutoCodeRover, and Agentless, further show that LLM-generated code becomes more useful when connected to repositories, execution feedback, and validation procedures~\cite{jimenez2024swebench,yang2024sweagent,zhang2024autocoderover,xia2025agentless}. Recent coding harnesses such as Codex and Claude Code make this capability visible in practical development workflows~\cite{openai2026codex,anthropic2026claudecode}. These works establish the coding side of the problem: LLMs can generate and repair executable programs when they are embedded in an environment that returns errors, tests, and other feedback.

Recent work also moves LLM agents from software repair toward data science and experiment workflows. Data Interpreter studies LLM-based data-science agents for long-term interconnected tasks, dynamic data changes, and verification beyond simple execution feedback~\cite{hong2024datainterpreter}. MLAgentBench and MLE-bench evaluate language agents on machine-learning experimentation and engineering tasks, where agents read and write files, execute code, inspect outputs, prepare datasets, and run experiments~\cite{huang2023mlagentbench,chan2024mlebench}. These studies make the evaluation target closer to automated experimentation: generated code must interact with data, produce intermediate outputs, and support a larger experimental objective rather than only pass a local programming test.

Networking and operations studies further show that LLMs can help interpret heterogeneous technical evidence. NetLLM, LLexus, and LogGPT apply LLMs to network reasoning, incident management, and log-anomaly analysis~\cite{wu2024netllm,lascasas2024llexus,han2023loggpt}. However, Starlink measurement analysis requires an additional acceptance layer beyond runnable code, successful task completion, or plausible diagnosis. A generated analysis script should be executable, compatible with the measurement schema, tied to real Starlink evidence, and able to produce the expected table, figure, benchmark, or case-study artifact. This leaves a workflow gap between LLM-based coding capability and validated, reusable measurement-analysis artifacts that can support automated experiments.

\section{Measurement-to-Artifact Problem Model}
\label{sec:formulation}

To characterize this workflow gap, we model Starlink measurement analysis as a measurement-to-artifact problem. The model defines five objects used by StarCodex: measurement batch, analysis state, analysis gap, structured coding task, and analysis artifact. These objects specify what is observed, what analysis is missing, and what reusable output should be produced.

\subsection{Measurement Batch}

Starlink measurements are represented in batches. A batch is the basic unit for observing new network evidence and deciding whether additional analysis is needed. It may correspond to a geographic region, a collection period, a source run, or a set of valid replay segments extracted from raw Starlink logs.

Formally, the $k$th measurement batch is denoted as
\begin{equation}
B_k =
\left(
\mathcal{T}_k,\bm{q}_k,\mathcal{E}_k
\right),
\label{eq:measurement-batch}
\end{equation}
where $\mathcal{T}_k=\{\tau_{k,1},\ldots,\tau_{k,n_k}\}$ is the set of valid Starlink measurement segments in the batch, $\bm{q}_k$ records data-quality and source metadata, and $\mathcal{E}_k$ records downstream evidence that has already been obtained. A segment $\tau_{k,j}$ contains the time-varying observations needed for later analysis, such as throughput, latency, handover-related information, and sampling coverage. The metadata $\bm{q}_k$ describes whether the batch contains missing fields, abnormal duration, schema changes, or other data-quality issues. The evidence set $\mathcal{E}_k$ may include severe-stall cases, prediction-overestimation cases, or other failures exposed by prediction or ABR replay.

This definition makes a batch an analyzable data unit with explicit source, quality, and downstream-evidence information. It records where the new data come from, whether the data are usable, and whether they already expose downstream prediction or playback risk.

\subsection{Analysis State}

The analysis state summarizes what is already known about the measurement corpus. It records the current status of data quality, network conditions, extracted features, failure evidence, experiment outputs, and accepted reusable artifacts. After processing batch $B_k$, the analysis state is written as
\begin{equation}
S_k =
\left(
S_k^{\mathrm{data}},
S_k^{\mathrm{profile}},
S_k^{\mathrm{feature}},
S_k^{\mathrm{failure}},
S_k^{\mathrm{exp}},
\mathcal{A}_k
\right),
\label{eq:analysis-state}
\end{equation}
where each term has a concrete role in gap detection and artifact reuse. Table~\ref{tab:analysis-state} maps the symbols in \eqref{eq:analysis-state} to the stored information. The first five components summarize the current measurement corpus from different analysis views, while $\mathcal{A}_k$ is the Artifact Memory that stores accepted artifacts for later batches and experiments.

\begin{table}[!t]
\centering
\caption{Components of the analysis state.}
\label{tab:analysis-state}
\footnotesize
\setlength{\tabcolsep}{2pt}
\renewcommand{\arraystretch}{1.12}
\begin{tabular}{@{}p{0.18\linewidth}p{0.25\linewidth}p{0.49\linewidth}@{}}
\toprule
\textbf{Symbol} & \textbf{Component} & \textbf{Stored Information} \\
\midrule
$S_k^{\mathrm{data}}$ & Data state & Valid and invalid segments, missing fields, schema version, and data-quality reports. \\
$S_k^{\mathrm{profile}}$ & Profile state & Known network states, candidate profiles, hard-segment distributions, and profile membership tables. \\
$S_k^{\mathrm{feature}}$ & Feature state & Existing measurement features, candidate features, and feature tables. \\
$S_k^{\mathrm{failure}}$ & Failure state & Severe-session cases, prediction-overestimation cases, unexplained failures, and failure reports. \\
$S_k^{\mathrm{exp}}$ & Experiment state & Generated or missing tables, figures, case studies, and experiment reports. \\
$\mathcal{A}_k$ & Artifact Memory & Accepted artifacts that can be reused by later batches or experiments. \\
\bottomrule
\end{tabular}
\end{table}

The analysis state provides context for subsequent coding tasks. For example, a schema change recorded in $S_k^{\mathrm{data}}$ may trigger a data-audit artifact, while hard cases stored in $S_k^{\mathrm{failure}}$ but not explained by $S_k^{\mathrm{profile}}$ may trigger a profile-mining or feature-synthesis artifact. Thus, the state components in Table~\ref{tab:analysis-state} serve as the trigger source for analysis gaps exposed by the next batch.

\subsection{Analysis Gap}

An \emph{analysis gap} is a missing analysis requirement exposed by new measurements relative to the current analysis state. It describes what the current analysis process cannot yet explain, check, or produce for the newly arrived batch. In practice, a new batch may reveal an unaudited data issue, an uncovered access condition, an unexplained failure pattern, or a missing experiment product.

The gap set $\mathcal{G}_k$ is organized according to the state components in \eqref{eq:analysis-state}:
\begin{equation}
\mathcal{G}_k
=
\mathcal{G}_k^{\mathrm{data}}
\cup
\mathcal{G}_k^{\mathrm{profile}}
\cup
\mathcal{G}_k^{\mathrm{feature}}
\cup
\mathcal{G}_k^{\mathrm{failure}}
\cup
\mathcal{G}_k^{\mathrm{exp}} .
\label{eq:gap-decomposition}
\end{equation}
Each subset in \eqref{eq:gap-decomposition} corresponds to a different kind of missing analysis. Table~\ref{tab:analysis-gap} gives typical examples of the gaps.

\begin{table}[!t]
\centering
\caption{Analysis gap categories and examples.}
\label{tab:analysis-gap}
\footnotesize
\setlength{\tabcolsep}{3pt}
\renewcommand{\arraystretch}{1.06}
\begin{tabular}{@{}p{0.18\linewidth}p{0.28\linewidth}p{0.40\linewidth}@{}}
\toprule
\textbf{Gap Subset} & \textbf{Meaning} & \textbf{Example Gap} \\
\midrule
$\mathcal{G}_k^{\mathrm{data}}$ & Missing data-quality, field, format, or valid-segment analysis. & A new batch contains missing throughput fields, invalid timestamps, or short segments that have not been audited. \\
$\mathcal{G}_k^{\mathrm{profile}}$ & Missing characterization of newly exposed network states or hard profiles. & A system-risk segment is not covered by the current low-tail, high-volatility, high-latency, or handover-heavy labels. \\
$\mathcal{G}_k^{\mathrm{feature}}$ & Missing measurement features needed to explain a risk or failure pattern. & Existing features cannot distinguish a prediction-overestimation window from a normal high-throughput window. \\
$\mathcal{G}_k^{\mathrm{failure}}$ & Missing diagnosis for prediction or ABR replay failures. & A segment causes severe rebuffering or prediction overestimation, but the failure evidence has not been summarized. \\
$\mathcal{G}_k^{\mathrm{exp}}$ & Missing experiment output needed to support a later analysis. & The current experiment lacks a benchmark table, comparison figure, or representative replay case. \\
\bottomrule
\end{tabular}
\end{table}

Thus, $\mathcal{G}_k$ turns heterogeneous measurement issues into explicit analysis targets. This representation keeps data quality, profile coverage, feature sufficiency, failure explanation, and experiment outputs within the same problem model.

\subsection{Structured Coding Task}

An analysis gap identifies a missing analysis, but it is not yet a concrete request that a coding harness can execute. A \emph{structured coding task} turns such a gap into an actionable specification: it names the missing analysis, provides the relevant measurement and state context, describes the expected artifact, and records how the generated output will be checked. Formally, the task for gap $g\in\mathcal{G}_k$ is written as
\begin{equation}
T_{k,g}
=
\left(
g,
\mathcal{C}_{k,g},
\mathcal{Y}_{k,g},
\mathcal{V}_{k,g}
\right),
\label{eq:structured-coding-task}
\end{equation}
where $\mathcal{C}_{k,g}$ is the compact context for the gap, $\mathcal{Y}_{k,g}$ specifies the expected artifact type and output form, and $\mathcal{V}_{k,g}$ specifies the validation requirements. 

\subsection{Analysis Artifact}

An analysis artifact is the output unit generated and checked in the measurement-to-artifact workflow. It is a reusable analysis product for a specific gap, such as a data auditor, profile miner, feature extractor, failure diagnostic script, benchmark construction script, figure script, or case-study selector. We represent an artifact as
\begin{equation}
a =
\left(
\mathrm{type},
g,
\mathcal{I},
\mathcal{O},
\pi,
r
\right),
\label{eq:analysis-artifact}
\end{equation}
where $\mathrm{type}$ is the artifact type, $g\in\mathcal{G}_k$ is the gap that the artifact addresses, $\mathcal{I}$ and $\mathcal{O}$ are the input and output constraints, $\pi$ is the generated executable program or analysis procedure, and $r$ is the resulting report, table, figure, or case list. Table~\ref{tab:artifact-types} gives examples of artifact types and their associated products. 

\begin{table}[!t]
\centering
\caption{Artifact types in the measurement-to-artifact model.}
\label{tab:artifact-types}
\footnotesize
\setlength{\tabcolsep}{3pt}
\renewcommand{\arraystretch}{1.06}
\begin{tabular}{@{}p{0.30\linewidth}p{0.27\linewidth}p{0.31\linewidth}@{}}
\toprule
\textbf{Artifact Type} & \textbf{Related Gap} & \textbf{Main Product} \\
\midrule
Data-audit artifact & Data gap & Valid-segment inventory and data-quality report. \\
Feature-extraction artifact & Feature gap & Per-segment feature table linking measurement, prediction, and replay evidence. \\
Hard-case discovery artifact & Profile or failure gap & Selected hard segments with replay and prediction evidence. \\
Benchmark construction artifact & Experiment gap & Reusable hard-case benchmark table and manifest. \\
Prediction-risk analysis artifact & Feature or failure gap & Risk-window table for prediction overestimation. \\
ABR replay artifact & Failure or experiment gap & Replay outcome table, QoE--risk figure, and case-level replay view. \\
\bottomrule
\end{tabular}
\end{table}

The analysis artifact is the unit through which generated code becomes reusable experiment material. Once accepted, an artifact can be executed, inspected, and reused by later analysis steps. This makes artifact quality the basis for evaluating whether dynamic coding contributes to experiment automation.

\section{StarCodex Framework}
\label{sec:starcodex-framework}

StarCodex can be described as a workflow from new data to accepted artifacts. The workflow detects analysis gaps, builds structured coding tasks, invokes Codex Dynamic Coding, validates candidate artifacts, returns repair feedback when needed, and updates Artifact Memory and the analysis state. 

\begin{figure}[!t]
  \centering
  \includegraphics[width=\linewidth]{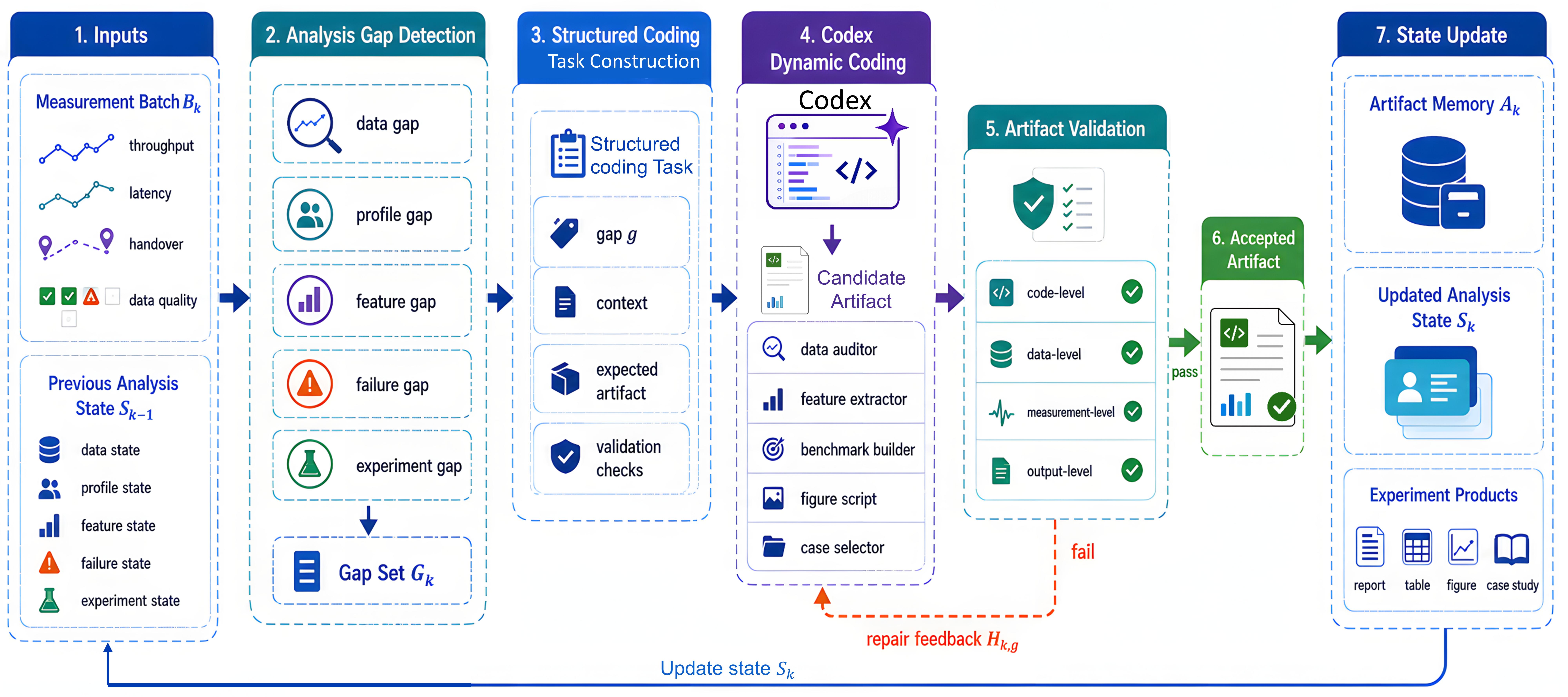}
    \caption{Overall workflow of StarCodex.}
  \label{fig:starcodex-framework}
\end{figure}

\subsection{Framework Structure}

The framework structure turns the measurement-to-artifact objective into an executable process. StarCodex is state-driven: the current analysis state determines which gaps are exposed, the gaps determine which coding tasks are constructed, validation determines which candidate artifacts are accepted, and accepted artifacts update the next analysis state. Fig.~\ref{fig:starcodex-framework} shows the overall workflow.

For batch $B_k$, the workflow can be summarized as
\begin{equation}
\left(
B_k,S_{k-1}
\right)
\rightarrow
\mathcal{G}_k
\rightarrow
\mathcal{Q}_k
\rightarrow
\widetilde{\mathcal{A}}_k
\rightarrow
\mathcal{A}_k^{\mathrm{acc}}
\rightarrow
S_k ,
\label{eq:starcodex-workflow}
\end{equation}
where $\mathcal{G}_k$ is the triggered analysis-gap set, $\mathcal{Q}_k=\{T_{k,g}:g\in\mathcal{G}_k\ \text{is selected}\}$ is the set of structured coding tasks constructed from those gaps, $\widetilde{\mathcal{A}}_k$ is the set of candidate artifacts generated by Codex, $\mathcal{A}_k^{\mathrm{acc}}$ is the subset that passes validation, and $S_k$ is the updated analysis state. This workflow is repeated as new measurement batches arrive, so later analysis can reuse previously accepted artifacts.

\subsection{Analysis Gap Detection}
\label{sec:gap-detection}

Analysis gap detection identifies what the current measurement batch still lacks before StarCodex starts coding. A newly arrived batch may contain invalid data fields, unfamiliar access behavior, unexplained failure evidence, or missing experiment products. The detector compares this new evidence with the current analysis state and the reusable artifacts already available, so that coding effort is assigned to missing analyses.

The output of this step is the active gap set for batch $B_k$. Formally, the detector maps the current batch, previous state, and Artifact Memory into
\begin{equation}
\mathcal{G}_k
=
\operatorname{DetectGap}
\left(
B_k,
S_{k-1},
\mathcal{A}_{k-1}
\right).
\label{eq:gap-detection}
\end{equation}
In practice, data gaps are triggered by missing fields, schema changes, invalid segments, or unresolved data-quality reports. Profile gaps are triggered when newly observed hard segments do not match the known profile state. Feature gaps are triggered when existing features cannot explain a failure or prediction-risk pattern. Failure gaps are triggered by prediction or ABR replay evidence not represented in the failure state. Experiment gaps are triggered when later analysis requires a benchmark, figure, table, or case study that has not yet been generated.

\subsection{Structured Coding Task Construction}

After gap detection, StarCodex converts each selected gap into the task object, so that the missing analysis can be passed to Codex Dynamic Coding. This step fills the task fields with the information needed by both stages. The context field summarizes the relevant batch and state evidence, the expected-output field specifies the artifact to be produced, and the validation field records the checks that will be applied after generation.

For a selected gap $g\in\mathcal{G}_k$, task construction can be written as
\begin{equation}
T_{k,g}
=
\operatorname{BuildTask}
\left(
g,
\operatorname{summary}(B_k,S_{k-1},\mathcal{A}_{k-1})
\right),
\label{eq:task-construction}
\end{equation}
where $\operatorname{BuildTask}(\cdot)$ instantiates the compact context $\mathcal{C}_{k,g}$, expected artifact description $\mathcal{Y}_{k,g}$, and validation requirements $\mathcal{V}_{k,g}$ in \eqref{eq:structured-coding-task}. The context may include the batch summary, relevant parts of the analysis state, failure evidence, accepted artifacts from Artifact Memory, and repair feedback from previous attempts.

\subsection{Codex Dynamic Coding}

Codex Dynamic Coding generates a candidate artifact from a structured task. It uses Codex to write or repair analysis code according to the task context and the previous validation feedback. This is the component that turns a selected analysis gap into an executable analysis artifact.

Given task $T_{k,g}$ and feedback history $\mathcal{H}_{k,g}$ for the same gap, the Codex-backed engine produces
\begin{equation}
\tilde a_{k,g}
=
C_{\theta}
\left(
T_{k,g},
\mathcal{H}_{k,g}
\right),
\label{eq:dynamic-coding-engine}
\end{equation}
where $C_{\theta}$ denotes the Codex-backed dynamic coding harness and $\tilde a_{k,g}$ is a candidate artifact. A candidate artifact normally contains an executable program, input and output descriptions, a run command, output paths, and a short explanation of the intended analysis.

\subsection{Artifact Validation and Repair Feedback}
\label{sec:artifact-validation}

\begin{figure}[!t]
  \centering
  \includegraphics[width=\linewidth]{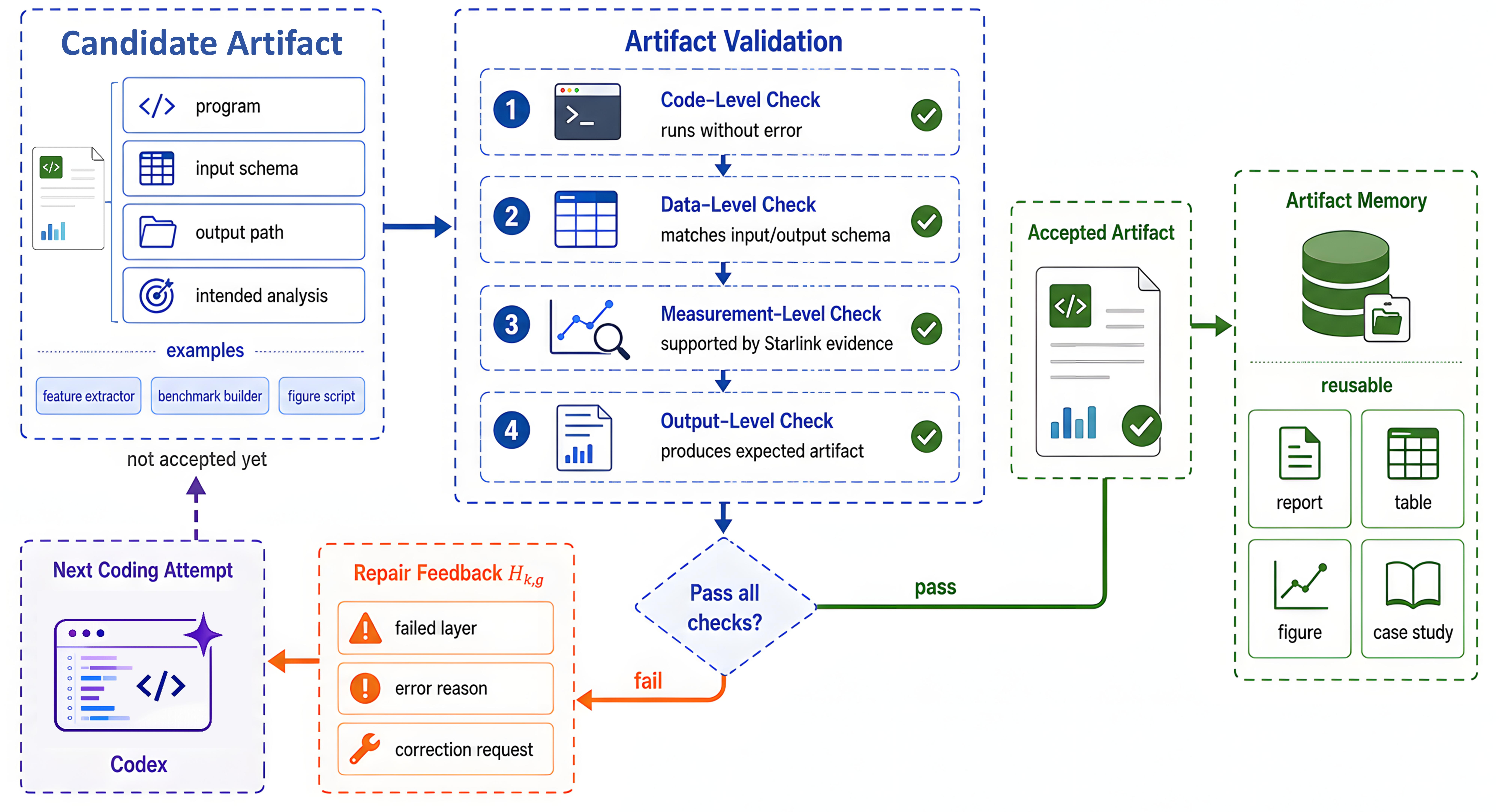}
  \caption{Artifact validation and repair feedback in StarCodex.}
  \label{fig:artifact-validation}
\end{figure}

Artifact validation determines whether a candidate artifact can become part of the accepted artifact set. StarCodex evaluates four validation components: code-level, data-level, measurement-level, and output-level validation. Fig.~\ref{fig:artifact-validation} illustrates the accept-or-repair mechanism for a candidate artifact.

For candidate artifact $\tilde a_{k,g}$, the acceptance indicator is shown as
\begin{equation}
\begin{aligned}
&\operatorname{Accept}(\tilde a_{k,g};B_k,S_{k-1})\\
&\quad =
V_{\mathrm{code}}(\tilde a_{k,g})
\wedge
V_{\mathrm{data}}(\tilde a_{k,g};B_k)\\
&\qquad \wedge
V_{\mathrm{meas}}(\tilde a_{k,g};B_k,S_{k-1})
\wedge
V_{\mathrm{out}}(\tilde a_{k,g}).
\end{aligned}
\label{eq:artifact-validation}
\end{equation}
The four terms correspond to the validation layers in Table~\ref{tab:validation-layers}.

\begin{table}[!t]
\centering
\caption{Validation layers used by StarCodex.}
\label{tab:validation-layers}
\footnotesize
\setlength{\tabcolsep}{3pt}
\renewcommand{\arraystretch}{1.06}
\begin{tabular}{@{}p{0.36\linewidth}p{0.58\linewidth}@{}}
\toprule
\textbf{Validation Layer} & \textbf{Purpose} \\
\midrule
Code-level validation & Check whether the generated program exists, compiles, executes, and is not a placeholder. \\
Data-level validation & Check whether the artifact declares and respects the expected input and output schemas. \\
Measurement-level validation & Check whether the artifact is tied to Starlink measurement evidence or replay-observed failure evidence. \\
Output-level validation & Check whether the expected report, table, figure, or case list exists and is non-empty. \\
\bottomrule
\end{tabular}
\end{table}

If candidate artifact $\tilde a_{k,g}$ passes the validation checks, it is added to the accepted set:
\begin{equation}
\mathcal{A}_k^{\mathrm{acc}}
\leftarrow
\mathcal{A}_k^{\mathrm{acc}}
\cup
\{\tilde a_{k,g}\}.
\label{eq:accepted-artifact-update}
\end{equation}
If it fails, StarCodex generates repair feedback. The feedback records the failed validation layer, the failure reason, and the required correction, such as schema mismatch, missing output file, empty report, unmatched measurement evidence, or unsupported figure output. This feedback is appended to $\mathcal{H}_{k,g}$ and becomes input to the next dynamic coding attempt for the same task.

\subsection{Accepted Artifact, Artifact Memory, and State Update}

Accepted artifacts are stored in Artifact Memory and used to update the analysis state. Artifact Memory is a lightweight registry that records each accepted artifact's type, path, source task, input and output constraints, and execution result. The analysis state then records the updates introduced by these artifacts to data quality, network states, features, failures, and experiment outputs.

At the end of batch $B_k$, Artifact Memory and the analysis state are updated as
\begin{equation}
\mathcal{A}_k
=
\mathcal{A}_{k-1}
\cup
\mathcal{A}_k^{\mathrm{acc}},
\label{eq:artifact-memory-update}
\end{equation}
and
\begin{equation}
S_k
=
\operatorname{Update}
\left(
S_{k-1},
B_k,
\mathcal{A}_k^{\mathrm{acc}}
\right).
\label{eq:state-update}
\end{equation}
For example, a data-audit artifact updates the data state, a profile-mining artifact updates the profile state, a feature-extraction artifact updates the feature state, a failure-diagnosis artifact updates the failure state, and a figure or report artifact updates the experiment state. This update rule allows StarCodex to accumulate analysis capability across batches.

\subsection{Algorithmic Summary}

StarCodex organizes dynamic coding as a stateful procedure over incoming measurement batches. Each batch may expose multiple analysis gaps, and each gap may require several coding and repair attempts before an artifact is accepted. Accepted artifacts are then stored in Artifact Memory and become part of the analysis state used for later batches. Algorithm~\ref{alg:starcodex} presents this dynamic coding procedure.

\begin{algorithm}[!t]
\caption{StarCodex Dynamic Coding Harness}
\label{alg:starcodex}
\begin{algorithmic}[1]
\STATE \textbf{Input:} Measurement batches $\{B_k\}$, initial analysis state $S_0$, initial Artifact Memory $\mathcal{A}_0$
\STATE \textbf{Parameters:} Maximum coding tasks per batch $M$, maximum repair rounds $R$
\FOR{each measurement batch $B_k$}
    \STATE Read $B_k$, previous state $S_{k-1}$, and Artifact Memory $\mathcal{A}_{k-1}$
    \STATE Detect analysis gaps $\mathcal{G}_k$ using \eqref{eq:gap-detection}
    \STATE Select at most $M$ gaps from $\mathcal{G}_k$
    \STATE Initialize $\mathcal{A}_k^{\mathrm{acc}}\leftarrow\emptyset$
    \FOR{each selected gap $g$}
        \STATE Construct coding task $T_{k,g}$ using \eqref{eq:task-construction}
        \STATE Initialize repair feedback history $\mathcal{H}_{k,g}\leftarrow\emptyset$
        \FOR{$r=1$ to $R$}
            \STATE Generate candidate artifact $\tilde a_{k,g}$ using \eqref{eq:dynamic-coding-engine}
            \STATE Evaluate $\operatorname{Accept}(\tilde a_{k,g};B_k,S_{k-1})$ using \eqref{eq:artifact-validation}
            \IF{$\operatorname{Accept}(\tilde a_{k,g};B_k,S_{k-1})=1$}
                \STATE Add $\tilde a_{k,g}$ to $\mathcal{A}_k^{\mathrm{acc}}$
                \STATE \textbf{break}
            \ELSE
                \STATE Append validation feedback to $\mathcal{H}_{k,g}$
            \ENDIF
        \ENDFOR
    \ENDFOR
    \STATE Update Artifact Memory using \eqref{eq:artifact-memory-update}
    \STATE Update analysis state $S_k$ using \eqref{eq:state-update}
    \STATE Export batch-level StarCodex report
\ENDFOR
\STATE \textbf{Output:} Accepted artifacts $\mathcal{A}$, analysis states $\{S_k\}$, and artifact results $\{r\}$
\end{algorithmic}
\end{algorithm}

\section{Experimental Evaluation}
\label{sec:evaluation}

\subsection{Experimental Setup}
\label{sec:experimental-setting}

\subsubsection{Starlink Measurement Data and Valid Replay Segments}
The experiments use real Starlink measurement data~\cite{liu2025starnet}. The measurements contain downlink throughput, latency, and connected-satellite records collected across different regions and time periods, and expose lower-tail capacity drops, throughput variation, elevated latency, and handover-related changes. We resample each source run at 1~Hz and divide it into non-overlapping 600-s segments. A segment is retained when throughput coverage is at least 95\%, latency coverage is at least 80\%, active-throughput coverage is at least 95\%, and the longest throughput and latency gaps do not exceed 5~s and 15~s, respectively.

Table~\ref{tab:starcodex-trace-inventory} summarizes the resulting corpus. The regional tags US, OSN, and VIC refer to Chicago in the United States, Osnabr\"uck in Germany, and Victoria in Canada. P05 and P50 denote the segment-level 5th-percentile and median throughput, and Lat. P95 denotes the 95th-percentile latency; the table reports regional averages of these segment-level statistics. Handover Changes is the regional mean number of timestamp-aligned connected-satellite changes per segment. Severe reports the percentage of segments for which the BOLA or RobustMPC screening replay exceeds 10~s of cumulative rebuffering, whereas Overest. reports the percentage containing a prediction-overestimation event.

\begin{table*}[!t]
\centering
\caption{Starlink measurement inventory.}
\label{tab:starcodex-trace-inventory}
\footnotesize
\setlength{\tabcolsep}{3.5pt}
\renewcommand{\arraystretch}{1.08}
\begin{tabular}{@{}lrrrrrrrrr@{}}
\toprule
\textbf{Region} & \textbf{Segments} & \textbf{Source} &
\multicolumn{3}{c}{\textbf{Throughput (Mbps)}} &
\textbf{Lat.} & \textbf{Handover} &
\multicolumn{2}{c}{\textbf{Failure Evidence (\%)}} \\
\cmidrule(lr){4-6}\cmidrule(l){9-10}
 & & \textbf{Runs} & \textbf{Mean} & \textbf{P05} & \textbf{P50} &
\textbf{P95} & \textbf{Changes} & \textbf{Severe} & \textbf{Overest.} \\
\midrule
OSN & 1,017 & 77 & 216.2 & 122.4 & 219.4 & 61.6 & 27.4 & 0.10 & 7.67 \\
US & 1,940 & 16 & 210.4 & 94.4 & 216.4 & 44.0 & 23.2 & 0.36 & 19.59 \\
VIC & 241 & 140 & 171.7 & 78.7 & 169.7 & 60.1 & 31.3 & 0.00 & 11.62 \\
\midrule
Total & 3,198 & 233 & 209.3 & 102.1 & 213.8 & 50.8 & 25.2 & 0.25 & 15.20 \\
\bottomrule
\end{tabular}
\end{table*}

The 3,198 valid segments originate from 233 source runs. We partition complete source runs into discovery, calibration, and test splits, so no source run contributes segments to more than one split. This fixed partition supports profile-threshold fitting and prediction-risk analysis. Hard-case discovery and benchmark construction instead use four source-run-grouped outer folds spanning all 233 runs, with separate fit, feedback, acceptance, and outer-test sets inside each fold. Table~\ref{tab:starcodex-data-protocol} summarizes the common data protocol and its derived evaluation units.

\begin{table}[!t]
\centering
\caption{Experimental data protocol.}
\label{tab:starcodex-data-protocol}
\footnotesize
\setlength{\tabcolsep}{2.5pt}
\renewcommand{\arraystretch}{1.05}
\begin{tabular}{@{}>{\raggedright\arraybackslash}p{0.37\linewidth}>{\raggedright\arraybackslash}p{0.55\linewidth}@{}}
\toprule
\textbf{Protocol Item} & \textbf{Setting} \\
\midrule
Segment construction & Non-overlapping 600-s segments at 1~Hz \\
Coverage & Throughput $\geq95\%$; active throughput $\geq95\%$; latency $\geq80\%$ \\
Maximum missing gap & Throughput $\leq5$~s; latency $\leq15$~s \\
\addlinespace[1pt]
Valid / excluded segments & 3,198 / 2,744 \\
Valid source runs & 233 \\
Split segments & 1,973 / 310 / 915 (discovery / calibration / test) \\
Split source runs & 116 / 47 / 70 (discovery / calibration / test) \\
\addlinespace[1pt]
System-risk segments & 486 \\
Discovery targets & 56 uncovered system-risk segments \\
Prediction-risk windows & 172,692 \\
Test windows / positives & 49,410 / 360 \\
Prediction-window construction & 60-s history, 10-s horizon, 10-s stride \\
Positive overestimation rule & History-mean prediction $>1.5\times$ subsequent mean capacity, with subsequent mean capacity $<20$~Mbps \\
\bottomrule
\end{tabular}
\end{table}

\subsubsection{Measurement Profiles and Failure Evidence}
Each valid segment is annotated with measurement-profile labels and system-level failure evidence. The labels capture low-tail capacity, high throughput volatility, elevated latency, and frequent handovers; their thresholds are fitted on discovery source runs and then frozen. Failure evidence is obtained independently from prediction analysis and ABR screening replay. A segment belongs to the system-risk set when BOLA or RobustMPC produces more than 10~s of cumulative rebuffering, or when point-capacity prediction overestimates a subsequent low-capacity interval. The profile labels may overlap, and they are not themselves used as failure outcomes.

\begin{table}[!t]
\centering
\caption{Measurement-profile groups and failure evidence in valid replay segments.}
\label{tab:profile-labels}
\footnotesize
\setlength{\tabcolsep}{3pt}
\renewcommand{\arraystretch}{1.06}
\begin{tabular}{@{}lrrr@{}}
\toprule
\textbf{Profile Group} & \textbf{Segments} & \textbf{Severe} & \textbf{Overest.} \\
 & & \textbf{Ratio} & \textbf{Ratio} \\
\midrule
Low-tail capacity & 601 & 0.013 & 0.646 \\
High volatility & 743 & 0.011 & 0.525 \\
High latency & 1,021 & 0.008 & 0.366 \\
Handover-heavy & 954 & 0.003 & 0.160 \\
Ordinary & 1,575 & 0.000 & 0.036 \\
\midrule
Uncovered system risk & 56 & 0.000 & 1.000 \\
\bottomrule
\end{tabular}
\end{table}

Table~\ref{tab:profile-labels} reports the number of segments associated with each profile group and the corresponding failure ratios. The first five rows describe measurement profiles; because the four risk-oriented labels may overlap, their counts are not intended to sum to the corpus size. The final row is the discovery target rather than an additional profile: it contains the 56 system-risk segments that fall outside all frozen risk-oriented profile rules. This separation lets the discovery experiment test whether an analysis artifact can find new risk evidence without using the target label as an input feature.

\subsubsection{Comparison Methods}
The comparison set is organized by analysis task because hard-case discovery, benchmark construction, prediction-risk analysis, and downstream replay produce different artifacts. Within each task, all methods receive the same eligible records, use the same source-run partitions, and operate under the same review, selection, or alert budget.

For hard-case discovery, Random samples from the complete eligible pool, whereas Profile-complement random samples only from segments outside the frozen profile-rule coverage. Isolation Forest (IF) supplies an unsupervised anomaly ranking. Fixed Extra Trees and fixed Random Forest (RF) are supervised scorers fitted on the outer-training partition. The Registry incumbent is the strongest artifact selected from the registered supervised scorers on the source-run-disjoint inner partition. StarCodex receives the same training-side evidence and registry tools, but Codex may generate or repair the executable scorer that is subsequently selected by the Harness. The random baselines are reported as repeated-sampling expectations.

Benchmark construction reuses these cross-fitted discovery scores under an equal 512-segment budget. Registry + IF combines the Registry-incumbent score with the IF anomaly score. StarCodex similarly combines its accepted discovery score with anomaly evidence. In both cases, the fusion weight is selected on the acceptance partition before the outer-test benchmark is constructed.

Prediction-risk analysis uses five task-specific baselines. Static expert applies a fixed weighted rule to recent throughput drop, history coefficient of variation, and lower-tail throughput. Point threshold ranks windows by the 60-s history-mean capacity. IF is fitted without labels on discovery and calibration features, whereas supervised RF uses the corresponding overestimation labels. Random gives a budget-matched reference ordering. StarCodex generates a scorer from discovery evidence, and the Harness commits it through the source-run-disjoint calibration gate. Every method receives the same history-derived measurements and contemporaneously available region indicators; future-capacity and label fields are excluded from scorer inputs. Evaluation uses the same source-run-disjoint test windows and matched alert budgets.

The downstream experiment compares the StarCodex benchmark with an equal-size reference set matched by region and measurement profile. BOLA, RobustMPC, Pensieve, and SafeSABR are replayed on both segment sets as downstream controllers rather than treated as analysis baselines. The capability ablation then compares the Registry artifact, the Initial artifact before validation-guided repair, the accepted artifact without risk-aware benchmark selection, and Full StarCodex. Here, \emph{Initial} denotes the first candidate produced within a StarCodex run, and \emph{Repaired} denotes the candidate revised from Harness feedback. Either becomes an accepted artifact after passing the Harness gate. The case study follows these artifact states within one StarCodex run; they describe the artifact lifecycle rather than additional comparison methods.

Codex serves as the dynamic coding backend that turns a structured task into an executable analysis artifact. The final discovery experiment used Codex \texttt{codex-cli} version 0.144.2 with the GPT-5.5 model and extra-high reasoning effort (\texttt{xhigh}). A generated Python artifact must pass code-level, data-level, measurement-level, and output-level validation; the Harness commits it only when it also satisfies the task-specific performance gate.

\subsubsection{Evaluation Metrics}
The evaluation metrics follow the output of each analysis task. For hard-case discovery, let $\mathcal{T}$ be the target cases in the outer-test partitions and $\mathcal{S}_m(q)$ be the top-ranked segments reviewed by method $m$ under budget $q$. An equal integer budget $\lceil qN/F\rceil$ is applied independently to each of the $F$ outer folds, where $N$ is the complete outer-test population. The target-hit count, discovery precision, and discovery recall are
\begin{equation}
\begin{aligned}
H_{\mathrm{disc}}(m,q)
&=|\mathcal{S}_m(q)\cap\mathcal{T}|,\\
\mathrm{Prec}@q(m)
&=\frac{H_{\mathrm{disc}}(m,q)}{|\mathcal{S}_m(q)|},\\
\mathrm{Rec}@q(m)
&=\frac{H_{\mathrm{disc}}(m,q)}{|\mathcal{T}|}.
\end{aligned}
\label{eq:true-discovery-metrics}
\end{equation}
Average precision (AP) is computed within each outer fold and aggregated using the number of positive targets in that fold as the weight; this avoids comparing uncalibrated score scales across folds. Source-run bootstrap intervals resample source runs within each outer fold while preserving the per-fold review budget. AP is reported for methods that produce a complete ranking. For the random methods, fixed-budget hit, precision, and recall values are expectations over repeated selections.

For benchmark construction, let $\mathcal{S}_m\subseteq\mathcal{D}_{\mathrm{disc}}$ be the fixed-size benchmark selected from the discovery-eligible set. A segment belongs to the system-risk set $\mathcal{R}$ when the BOLA or RobustMPC screening replay produces more than 10 s of cumulative rebuffering, or when the segment contains a prediction-overestimation event. The target set $\mathcal{T}\subseteq\mathcal{R}$ contains system-risk segments outside the frozen profile-rule coverage. We report uncovered-risk hits, system-risk hits, uncovered-risk recall, and system-risk rate:
\begin{equation}
\begin{aligned}
H_{\mathrm{unc}}(m)
&=|\mathcal{S}_m\cap\mathcal{T}|,\\
H_{\mathrm{risk}}(m)
&=|\mathcal{S}_m\cap\mathcal{R}|,\\
\mathrm{Rec}_{\mathrm{unc}}(m)
&=\frac{H_{\mathrm{unc}}(m)}{|\mathcal{T}|},\\
\mathrm{Rate}_{\mathrm{risk}}(m)
&=\frac{H_{\mathrm{risk}}(m)}{|\mathcal{S}_m|}.
\end{aligned}
\label{eq:benchmark-risk-metrics}
\end{equation}
The number of represented source runs measures the measurement-source breadth of the selected benchmark. The complete set $\mathcal{C}_{\mathrm{all}}$ contains four failure types: BOLA-severe replay, RobustMPC-severe replay, prediction overestimation, and abrupt capacity drop. Let $\mathcal{C}(\mathcal{S}_m)$ be the subset represented by the selected benchmark. Failure-type coverage (FTC) is
\begin{equation}
\mathrm{FTC}(m)
=
\frac{|\mathcal{C}(\mathcal{S}_m)|}{|\mathcal{C}_{\mathrm{all}}|}.
\label{eq:failure-type-coverage}
\end{equation}
FTC equals one when the benchmark represents every evaluated failure type.

For prediction-risk analysis, let $\mathcal{W}$ be the source-run-disjoint test-window set, $\widehat{\mathcal{W}}_m(q)$ be the windows selected under alert budget $q$, and $\mathcal{W}_{\mathrm{oe}}$ be the overestimation-positive windows. The overestimation hit count, recall, and precision are
\begin{equation}
\begin{aligned}
H_{\mathrm{oe}}(m,q)
&=|\widehat{\mathcal{W}}_m(q)\cap\mathcal{W}_{\mathrm{oe}}|,\\
\mathrm{Rec}_{\mathrm{oe}}(m,q)
&=\frac{H_{\mathrm{oe}}(m,q)}{|\mathcal{W}_{\mathrm{oe}}|},\\
\mathrm{Prec}_{\mathrm{oe}}(m,q)
&=\frac{H_{\mathrm{oe}}(m,q)}{|\widehat{\mathcal{W}}_m(q)|}.
\end{aligned}
\label{eq:prediction-risk-metrics}
\end{equation}
To capture the magnitude of the recognized prediction error, let $e(w)$ be the positive difference between point prediction and subsequently measured mean capacity in the positive window $w$. Overprediction-exposure coverage is
\begin{equation}
\mathrm{Cov}_{\mathrm{exp}}(m,q)
=
\frac{\sum_{w\in\widehat{\mathcal{W}}_m(q)\cap\mathcal{W}_{\mathrm{oe}}}e(w)}
{\sum_{w\in\mathcal{W}_{\mathrm{oe}}}e(w)}.
\label{eq:overprediction-exposure}
\end{equation}
This metric weights detected windows by the amount of capacity overestimation, rather than treating all positive windows as equally consequential. AP evaluates the complete ranking. Alert sets contain the top $\operatorname{round}(q|\mathcal{W}|)$ windows and are evaluated at 1\%, 2.5\%, 5\%, and 10\% budgets, with $q=5\%$ used for the exact-value table.

For downstream ABR replay, each controller--segment pair produces one replay session. Let $C$ be a controller, $B(C,\tau)$ be cumulative rebuffering on segment $\tau$, and $\bar r(C,\tau)$ be average bitrate. We use $b_{\mathrm{sev}}=10$ s for severe-session counting and $\alpha=0.05$ for worst-tail rebuffering. For a segment set $\mathcal{B}$,
\begin{equation}
\begin{aligned}
\bar r(C,\mathcal{B})
&=
\frac{1}{|\mathcal{B}|}
\sum_{\tau\in\mathcal{B}}\bar r(C,\tau),\\
\bar B(C,\mathcal{B})
&=
\frac{1}{|\mathcal{B}|}
\sum_{\tau\in\mathcal{B}} B(C,\tau),\\
R_{\mathrm{sev}}(C,\mathcal{B})
&=
\frac{1}{|\mathcal{B}|}
\sum_{\tau\in\mathcal{B}}
\mathbf{1}\{B(C,\tau)>b_{\mathrm{sev}}\}.
\end{aligned}
\label{eq:abr-replay-metrics}
\end{equation}
Worst-tail rebuffering is
\begin{equation}
T_{\alpha}(C,\mathcal{B})
=
\frac{1}{|\operatorname{Tail}_{\alpha}(C,\mathcal{B})|}
\sum_{\tau\in \operatorname{Tail}_{\alpha}(C,\mathcal{B})}
B(C,\tau),
\label{eq:abr-tail-metric}
\end{equation}
where $\operatorname{Tail}_{\alpha}(C,\mathcal{B})$ contains the $\lceil\alpha|\mathcal{B}|\rceil$ replay sessions with the largest cumulative rebuffering.

The capability ablation reuses the discovery and benchmark metrics above. Accepted Folds counts the outer folds in which a dynamically coded artifact passes validation and the Harness performance gate and is committed instead of the Registry artifact. The end-to-end case study uses the same fold-local discovery metrics and reports the artifact rank transition together with prediction and ABR replay evidence for the admitted segment. Its Mean Buffer value is the arithmetic mean of the per-chunk playback-buffer samples in that replay session.

\subsection{Hard-Case Discovery}
\label{sec:hard-case-discovery}
This experiment evaluates StarCodex for discovering system-risk Starlink segments that fall outside the frozen profile-rule coverage under a limited review budget. The objective is to determine whether dynamically generated and repaired discovery artifacts can concentrate newly uncovered risk evidence as effectively as strong fixed analysis models.

\begin{figure}[!t]
  \centering
  \includegraphics[width=\linewidth]{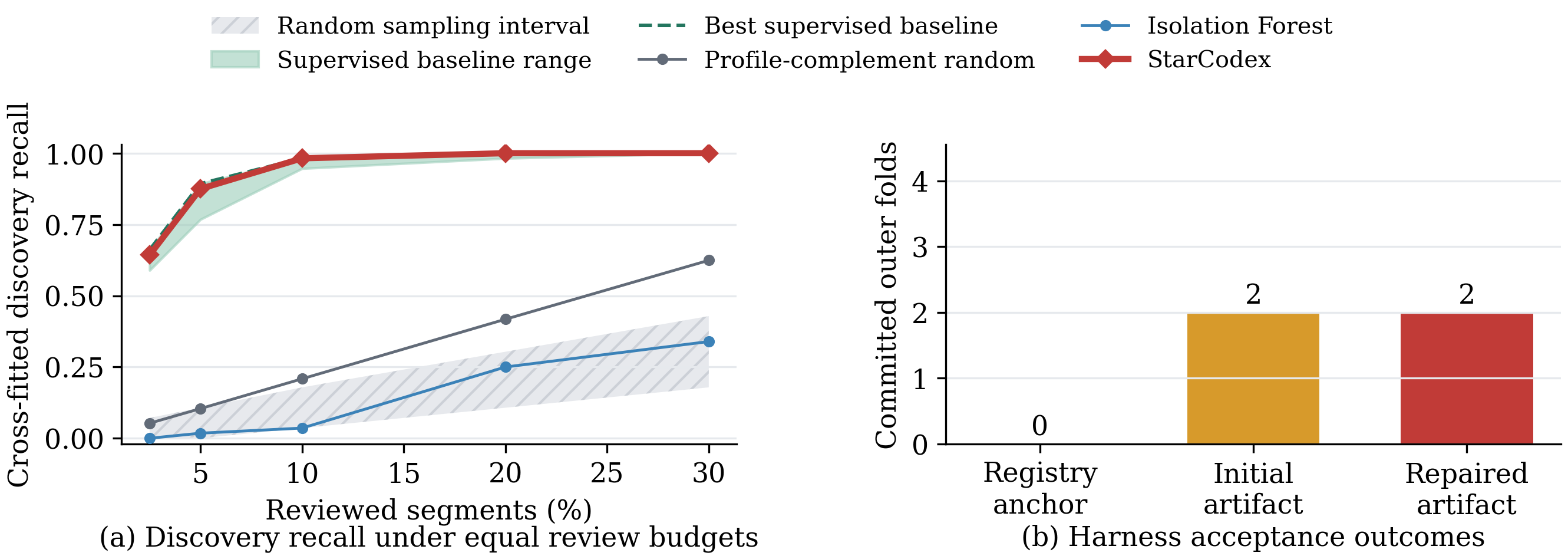}
  \caption{Uncovered system-risk recall across review budgets and Harness-selected artifact sources.}
  \label{fig:hard-case-discovery}
\end{figure}

The experiment uses the four source-run-grouped outer folds defined in Section~\ref{sec:experimental-setting}. Within each fold, fit runs support artifact construction, feedback runs support repair, acceptance runs determine commitment, and outer-test runs provide the final ranking. Each method ranks the outer-test segments independently, and the same integer review budget is applied to every ranking. The main operating point reviews the top 5\% per fold, corresponding to 160 segments in total; Fig.~\ref{fig:hard-case-discovery}(a) additionally varies this budget from 2\% to 30\%. Discovery quality is reported using the target-hit, precision, recall, and average-precision metrics defined in \eqref{eq:true-discovery-metrics}. Fig.~\ref{fig:hard-case-discovery}(b) records whether the Harness commits the Initial or Repaired artifact in each outer fold.

\begin{table}[!t]
\centering
\caption{Uncovered system-risk discovery with 160 reviewed segments (5\% budget); random rows report repeated-sampling means.}
\label{tab:hard-case-discovery}
\footnotesize
\setlength{\tabcolsep}{2.5pt}
\renewcommand{\arraystretch}{1.06}
\resizebox{\linewidth}{!}{%
\begin{tabular}{lcccc}
\toprule
\textbf{Method} & \textbf{Hits} & \textbf{Prec.@5\%} & \textbf{Rec.@5\%} & \textbf{AP} \\
\midrule
Random & 2.83 & 0.018 & 0.051 & -- \\
Profile-complement random & 5.83 & 0.036 & 0.104 & -- \\
IF & 1 & 0.006 & 0.018 & 0.025 \\
Fixed Extra Trees & 43 & 0.269 & 0.768 & 0.609 \\
Fixed RF & 49 & 0.306 & 0.875 & 0.690 \\
Registry incumbent & \textbf{50} & \textbf{0.313} & \textbf{0.893} & 0.689 \\
StarCodex & 49 & 0.306 & 0.875 & \textbf{0.703} \\
\bottomrule
\end{tabular}
}
\end{table}

At the 5\% review budget, Table~\ref{tab:hard-case-discovery} shows that StarCodex finds 49 of the 56 targets, matching fixed RF and finding six more targets than fixed Extra Trees. The 95\% source-run bootstrap interval for the Recall@5\% difference between StarCodex and fixed RF is $[-0.055,\,0.054]$, so we interpret their operating-point recall as comparable. The Registry incumbent finds one additional target, but StarCodex produces the highest AP, 0.703, compared with 0.690 for fixed RF and 0.689 for the Registry incumbent. Thus, StarCodex preserves strong discovery at the operating budget while providing the best complete-ranking quality. By comparison, profile-complement and unrestricted random sampling produce only 5.83 and 2.83 expected hits, showing that restricting the candidate pool to segments outside existing profile coverage is not sufficient by itself.

Fig.~\ref{fig:hard-case-discovery}(a) provides the budget-level trend. StarCodex follows the upper boundary of the supervised baseline range from 2\% to 30\% review budgets and remains well above IF and both random baselines. Fig.~\ref{fig:hard-case-discovery}(b) further shows that the Harness commits the Initial artifact in two outer folds and the Repaired artifact in the other two; none of the folds falls back to the Registry anchor. Overall, StarCodex reaches strong supervised-level discovery of system-risk cases outside the existing profile coverage while dynamically generating and repairing the analysis logic.

\subsection{Benchmark Construction Artifact Quality}
\label{sec:benchmark-construction}
This experiment evaluates whether StarCodex can convert cross-fitted discovery evidence into a fixed-size benchmark that retains uncovered-risk cases while adding broader system-risk evidence. The two quantities are measured by uncovered-risk recall and system-risk rate in \eqref{eq:benchmark-risk-metrics}.

\begin{figure}[!t]
  \centering
  \includegraphics[width=\linewidth]{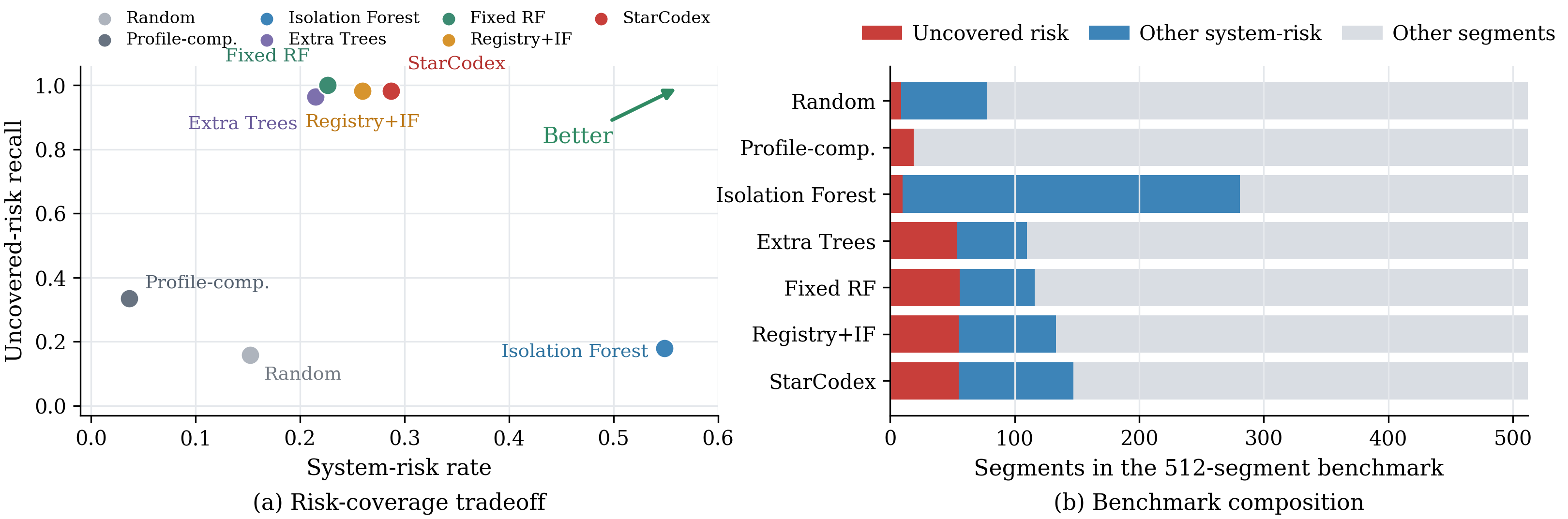}
  \caption{Risk coverage and composition of the 512-segment benchmarks.}
  \label{fig:benchmark-construction}
\end{figure}

Each method selects 128 segments independently from each outer-test fold, producing an equal 512-segment benchmark. Selection uses only the cross-fitted discovery and anomaly scores; uncovered-risk labels, system-risk labels, and failure indicators are read only after the benchmark has been fixed. For Registry + IF and StarCodex, the fusion weight is fixed on the source-run-disjoint acceptance partition before outer-test selection. Random results are reported as repeated-selection expectations. In Fig.~\ref{fig:benchmark-construction}, Profile-comp. abbreviates profile-complement random sampling. The figure reports all methods, while Table~\ref{tab:benchmark-construction} focuses on methods whose uncovered-risk recall exceeds 0.95.

\begin{table}[!t]
\centering
\caption{High-recall benchmark construction results at 512 selected segments.}
\label{tab:benchmark-construction}
\scriptsize
\setlength{\tabcolsep}{2pt}
\renewcommand{\arraystretch}{1.06}
\resizebox{\linewidth}{!}{%
\begin{tabular}{lccccc}
\toprule
\textbf{Method} & \textbf{Uncov. Hits} & \textbf{Uncov. Rec.} & \textbf{Risk Hits} & \textbf{Risk Rate} & \textbf{Source Runs} \\
\midrule
Fixed Extra Trees & 54 & 0.964 & 110 & 0.215 & 26 \\
Fixed RF & \textbf{56} & \textbf{1.000} & 116 & 0.227 & 30 \\
Registry + IF & 55 & 0.982 & 133 & 0.260 & 33 \\
StarCodex & 55 & 0.982 & \textbf{147} & \textbf{0.287} & \textbf{35} \\
\bottomrule
\end{tabular}
}
\end{table}

Fig.~\ref{fig:benchmark-construction}(a) shows the risk-coverage tradeoff. Fixed RF retains all 56 uncovered-risk cases, whereas StarCodex retains 55, giving uncovered-risk recall of 1.000 and 0.982, respectively. StarCodex places more of the fixed benchmark budget on system-risk evidence: its system-risk rate reaches 0.287, compared with 0.227 for fixed RF and 0.260 for Registry + IF. IF reaches a higher system-risk rate but recalls only 0.179 of the uncovered-risk cases, showing that anomaly density alone does not preserve the newly exposed cases.

Fig.~\ref{fig:benchmark-construction}(b) and Table~\ref{tab:benchmark-construction} show how this tradeoff changes benchmark composition. StarCodex selects 147 system-risk segments, including 55 uncovered-risk cases, whereas fixed RF selects 116 system-risk segments and Registry + IF selects 133. StarCodex therefore adds 31 system-risk segments over fixed RF, a 26.7\% increase, while missing one of the 56 uncovered-risk cases. It also preserves the same 0.982 uncovered-risk recall as Registry + IF while adding 14 system-risk segments and covering two additional source runs.

Under the same 512-segment budget, StarCodex thus preserves near-complete uncovered-risk coverage while constructing a benchmark with more system-risk evidence than the strong supervised and hybrid baselines. This result confirms that the accepted discovery artifact can be compiled into a broader benchmark while retaining almost all uncovered-risk cases that motivated its construction.

\subsection{Prediction-Risk Analysis Artifact Quality}
\label{sec:prediction-side-risk}
This experiment evaluates whether the prediction-risk artifact identifies windows in which point-capacity prediction substantially overestimates subsequently measured capacity. The evaluation measures both the coverage of these unsafe windows and the concentration of true overestimation evidence within a bounded alert set.

\begin{figure}[!t]
  \centering
  \includegraphics[width=\linewidth]{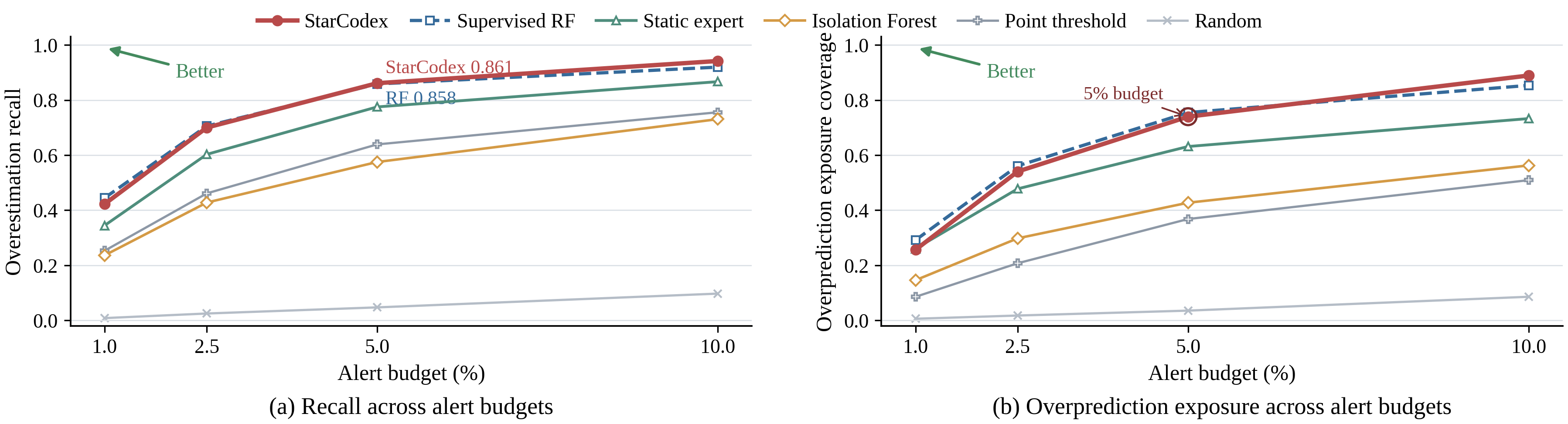}
  \caption{Prediction-overestimation recall and exposure coverage across matched alert budgets.}
  \label{fig:prediction-risk-recognition}
\end{figure}

Prediction-risk records are constructed from a 60-s history window, a 10-s prediction horizon, and a 10-s stride. A positive window is one in which the history-mean point prediction exceeds 1.5 times the subsequently measured mean capacity while that capacity is below 20~Mbps. We use the source-run-disjoint discovery, calibration, and test partitions summarized in Table~\ref{tab:starcodex-data-protocol}. All methods receive the same alert budget: Fig.~\ref{fig:prediction-risk-recognition} varies it from 1\% to 10\%, and Table~\ref{tab:prediction-risk-recognition} reports the main 5\% operating point using the metrics in \eqref{eq:prediction-risk-metrics} and \eqref{eq:overprediction-exposure}. In the table, OE denotes overestimation, Exposure denotes overprediction-exposure coverage, and Rec. and Prec. denote recall and precision.

\begin{table}[!t]
\centering
\caption{Prediction-risk analysis artifact quality under a 5\% alert budget.}
\label{tab:prediction-risk-recognition}
\footnotesize
\setlength{\tabcolsep}{2.2pt}
\renewcommand{\arraystretch}{1.06}
\resizebox{\linewidth}{!}{%
\begin{tabular}{lccccc}
\toprule
\textbf{Method} & \textbf{OE Hits} & \textbf{OE Rec.} & \textbf{OE Prec.} & \textbf{Exposure} & \textbf{AP} \\
\midrule
Random & 17 & 0.047 & 0.007 & 0.035 & 0.007 \\
Point threshold & 230 & 0.639 & 0.093 & 0.368 & 0.116 \\
Static expert & 279 & 0.775 & 0.113 & 0.631 & 0.199 \\
IF & 207 & 0.575 & 0.084 & 0.427 & 0.118 \\
Supervised RF & 309 & 0.858 & 0.125 & \textbf{0.754} & 0.288 \\
StarCodex & \textbf{310} & \textbf{0.861} & \textbf{0.126} & 0.739 & \textbf{0.291} \\
\bottomrule
\end{tabular}
}
\end{table}

Fig.~\ref{fig:prediction-risk-recognition}(a) shows that the StarCodex prediction-risk artifact remains close to the supervised RF across all alert budgets and attains slightly higher recall at the 5\% and 10\% operating points. At 5\%, Table~\ref{tab:prediction-risk-recognition} reports 310 detected overestimation windows for StarCodex and 309 for supervised RF, corresponding to recalls of 0.861 and 0.858. The StarCodex artifact also reaches a complete-ranking AP of 0.291, compared with 0.288 for supervised RF and 0.199 for Static expert.

Fig.~\ref{fig:prediction-risk-recognition}(b) shows that supervised RF attains an overprediction-exposure coverage of 0.754 at the 5\% budget, slightly above the 0.739 achieved by StarCodex. At the 10\% budget, the ordering reverses: StarCodex reaches 0.889 exposure coverage, compared with 0.853 for supervised RF. Table~\ref{tab:prediction-risk-recognition} also reports similar 5\% precision for StarCodex and supervised RF, at 0.126 and 0.125, respectively. At the main 5\% operating point, both methods outperform Static expert, Point threshold, IF, and Random in recall, precision, and exposure coverage. These results show that dynamic coding can produce an executable prediction-risk artifact competitive with a purpose-trained supervised model.

\subsection{Downstream ABR Replay Impact}
\label{sec:abr-replay-impact}
This experiment examines whether the benchmark produced by StarCodex exposes downstream differences in ABR playback behavior. Actual controller replay connects the discovered measurement evidence to delivered bitrate and session-level rebuffering risk.

\begin{figure}[!t]
  \centering
  \includegraphics[width=\linewidth]{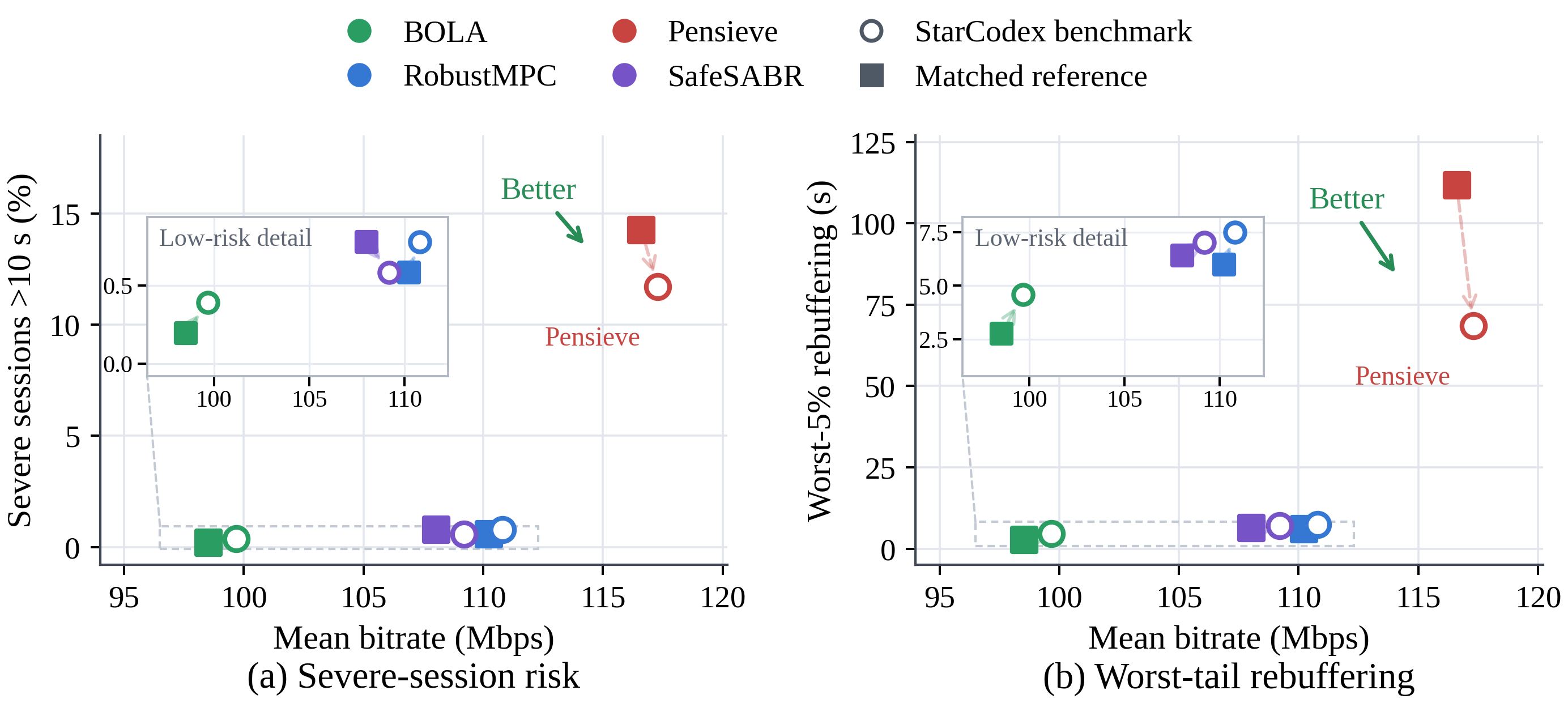}
  \caption{ABR quality--risk outcomes on the StarCodex benchmark and matched reference. The insets enlarge the low-risk region.}
  \label{fig:abr-replay-tradeoff}
\end{figure}

We replay BOLA~\cite{spiteri2020bola}, RobustMPC~\cite{yin2015control}, Pensieve~\cite{mao2017neural}, and SafeSABR~\cite{xie2026safesabr} on the 512 segments selected by StarCodex. The matched reference contains 512 different valid segments and reproduces the benchmark's regional and measurement-profile composition. Fig.~\ref{fig:abr-replay-tradeoff} shows the resulting quality--risk operating points. Open circles denote the StarCodex benchmark, filled squares denote the matched reference, and each dashed arrow points from the reference to the benchmark for the same controller. Table~\ref{tab:abr-replay-results} reports the corresponding replay outcomes.

\begin{table}[!t]
\centering
\caption{Downstream ABR replay outcomes on equal-size segment sets.}
\label{tab:abr-replay-results}
\scriptsize
\setlength{\tabcolsep}{2pt}
\renewcommand{\arraystretch}{1.06}
\resizebox{\linewidth}{!}{%
\begin{tabular}{@{}llrrrr@{}}
\toprule
\textbf{Set} & \textbf{Controller} & \textbf{Bitrate} & \textbf{Rebuf.} & \textbf{Worst-5\%} & \textbf{Severe} \\
 & & \textbf{(Mbps)} & \textbf{(s)} & \textbf{(s)} & \textbf{($>10$ s)} \\
\midrule
StarCodex & BOLA & 99.68 & 0.47 & 4.58 & 0.39\% \\
StarCodex & RobustMPC & 110.79 & 0.67 & 7.49 & 0.78\% \\
StarCodex & Pensieve & 117.29 & 5.42 & 68.51 & 11.72\% \\
StarCodex & SafeSABR & 109.20 & 0.62 & 7.02 & 0.59\% \\
Reference & BOLA & 98.53 & 0.39 & 2.78 & 0.20\% \\
Reference & RobustMPC & 110.24 & 0.57 & 5.98 & 0.59\% \\
Reference & Pensieve & 116.61 & 8.67 & 111.74 & 14.26\% \\
Reference & SafeSABR & 108.03 & 0.59 & 6.42 & 0.78\% \\
\bottomrule
\end{tabular}
}
\end{table}

The StarCodex benchmark produces a clear controller-level quality--risk separation. BOLA delivers 99.68 Mbps with 0.39\% severe sessions and 4.58 s worst-5\% rebuffering. RobustMPC and SafeSABR increase mean bitrate to 110.79 and 109.20 Mbps while keeping the severe-session ratio below 0.8\%. Pensieve reaches the highest mean bitrate, 117.29 Mbps, but its severe-session ratio rises to 11.72\%, and its worst-5\% rebuffering reaches 68.51 s. The discovered segments therefore expose distinct operating points rather than producing a bitrate-only controller ranking.

The matched-reference comparison further shows that the downstream effect is controller-specific. Moving from the matched reference to the StarCodex benchmark, BOLA worst-5\% rebuffering changes from 2.78 to 4.58 s, and RobustMPC changes from 5.98 to 7.49 s. SafeSABR worst-5\% rebuffering changes from 6.42 to 7.02 s, while its severe-session ratio changes from 0.78\% to 0.59\%. Pensieve follows a different direction: its severe-session ratio changes from 14.26\% to 11.72\%, and its worst-5\% rebuffering changes from 111.74 to 68.51 s. StarCodex converts measurement discoveries into executable playback evidence that reveals both the quality--risk separation and the controller specificity of downstream ABR behavior.

\subsection{Capability Ablation}
\label{sec:capability-ablation}
This experiment isolates how dynamic coding, validation-guided repair, and benchmark selection affect the artifacts produced by StarCodex.

All variants use the same source-run-grouped outer folds, 5\% discovery budget, and 512-segment benchmark budget. The progressive variants restore one capability at a time. The first removes dynamic coding and uses the stored Registry artifact. The second enables dynamic coding but retains the Initial artifact without validation-guided repair. The third adds repair but ranks benchmark segments directly by the accepted discovery score, without risk-aware benchmark selection. Full StarCodex combines the accepted discovery score with anomaly evidence using a fusion weight selected on the source-run-disjoint acceptance partition. Discovery and benchmark effects are measured with the same metrics as Sections~\ref{sec:hard-case-discovery} and~\ref{sec:benchmark-construction}; Accepted Folds records how many outer folds commit a dynamically coded artifact.

\begin{figure}[!t]
  \centering
  \includegraphics[width=\linewidth]{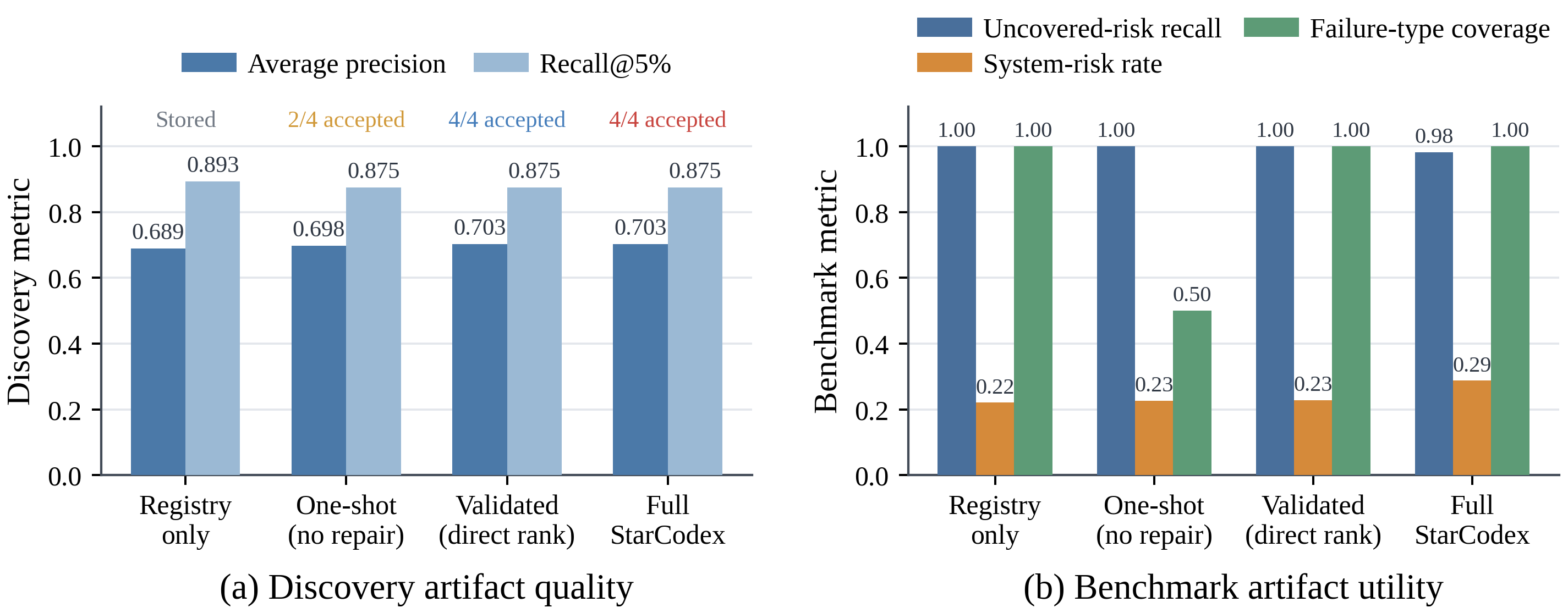}
  \caption{Same-pipeline capability ablation for discovery and benchmark construction.}
  \label{fig:capability-ablation}
\end{figure}

\begin{table}[!t]
\centering
\caption{Same-pipeline capability ablation of StarCodex.}
\label{tab:capability-ablation}
\scriptsize
\setlength{\tabcolsep}{2pt}
\renewcommand{\arraystretch}{1.06}
\resizebox{\linewidth}{!}{%
\begin{tabular}{@{}lccccccc@{}}
\toprule
\textbf{Variant} & \textbf{Acc.} & \textbf{AP} & \textbf{Rec.} & \textbf{Uncov.} & \textbf{Risk} & \textbf{Src.} & \textbf{FTC} \\
 & \textbf{Folds} & & \textbf{@5\%} & \textbf{Rec.} & \textbf{Rate} & \textbf{Runs} & \\
\midrule
w/o dynamic coding & -- & 0.689 & 0.893 & 1.000 & 0.221 & 29 & 1.000 \\
w/o validation-guided repair & 2/4 & 0.698 & 0.875 & 1.000 & 0.227 & 32 & 0.500 \\
w/o benchmark selection & 4/4 & 0.703 & 0.875 & 1.000 & 0.229 & 31 & 1.000 \\
Full StarCodex & \textbf{4/4} & \textbf{0.703} & 0.875 & 0.982 & \textbf{0.287} & \textbf{35} & \textbf{1.000} \\
\bottomrule
\end{tabular}
}
\end{table}

Fig.~\ref{fig:capability-ablation}(a) and Table~\ref{tab:capability-ablation} show the discovery-side effects. The stored Registry artifact reaches 0.689 AP and 0.893 Recall@5\%. The Initial artifact raises AP to 0.698, but dynamically coded artifacts are committed in only two of four outer folds. With validation-guided repair, a dynamically coded artifact is committed in all four folds: the Initial artifact in two folds and a feedback-repaired artifact in the other two. AP increases to 0.703, while Recall@5\% remains 0.875. The repair step therefore improves both artifact acceptance and complete-ranking quality without relying on a larger review budget.

Fig.~\ref{fig:capability-ablation}(b) shows the effect of benchmark selection. Direct ranking with the validated artifact retains all 56 uncovered-risk cases and yields a system-risk rate of 0.229 across 31 source runs. Full StarCodex retains 55 of the 56 cases, corresponding to 0.982 uncovered-risk recall, while increasing the system-risk rate to 0.287 and source-run coverage to 35. Both variants cover all failure types. Thus, benchmark selection adds broader system-risk and source-run evidence at the cost of one uncovered-risk case.

The ablation assigns a distinct role to each capability: dynamic coding improves AP over the stored Registry artifact, validation-guided repair increases the number of folds with an accepted artifact and further improves AP, and benchmark selection broadens the evidence represented by the final benchmark.

\subsection{End-to-End Artifact-Repair Case Study}
\label{sec:end-to-end-case-study}
This case study illustrates how validation-guided dynamic coding changes an executable discovery artifact and brings system-relevant measurement evidence into the benchmark. It follows one artifact from its initial validation result through repair, benchmark admission, prediction analysis, and downstream ABR replay.

The case is selected after the cross-fitted rankings have been fixed. Among outer folds in which StarCodex commits a repaired artifact, we identify segments ranked outside the per-fold benchmark cutoff of 128 by the initial artifact but inside the cutoff after repair. We then focus on a rank-crossing segment that produces severe replay outcomes for all four controllers. Artifact acceptance uses the source-run-disjoint acceptance partition, whereas the reported rank transition and controller replays use the corresponding outer-test fold. Here, \emph{Initial} denotes the pre-repair artifact within the same StarCodex run.

\begin{figure}[!t]
  \centering
  \includegraphics[width=\linewidth]{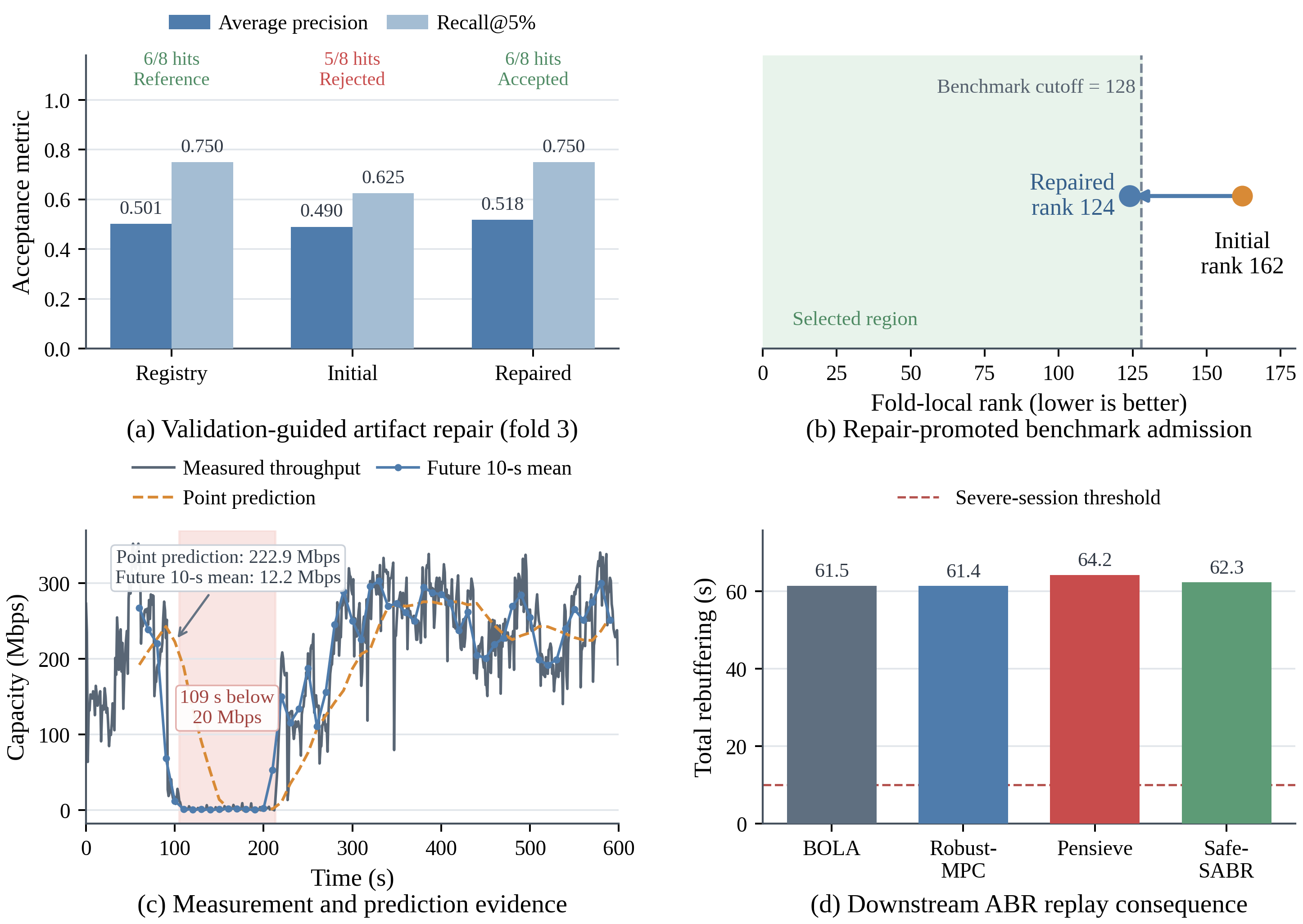}
  \caption{End-to-end case study of validation-guided artifact repair, benchmark admission, measurement evidence, and downstream ABR replay.}
  \label{fig:end-to-end-case-study}
\end{figure}

\begin{table}[!t]
\centering
\caption{Downstream replay outcomes on the admitted case segment.}
\label{tab:end-to-end-case-study}
\footnotesize
\setlength{\tabcolsep}{3.5pt}
\renewcommand{\arraystretch}{1.06}
\begin{tabular}{lccc}
\toprule
\textbf{Controller} & \textbf{Mean Bitrate} & \textbf{Mean Buffer} & \textbf{Total Rebuf.} \\
 & \textbf{(Mbps)} & \textbf{(s)} & \textbf{(s)} \\
\midrule
BOLA & 84.00 & \textbf{35.33} & 61.48 \\
RobustMPC & \textbf{100.48} & 26.66 & \textbf{61.41} \\
Pensieve & 98.17 & 23.98 & 64.22 \\
SafeSABR & 95.81 & 26.93 & 62.32 \\
\bottomrule
\end{tabular}
\end{table}

Fig.~\ref{fig:end-to-end-case-study}(a) shows the artifact-level repair in outer fold 3. The Registry artifact obtains 0.501 AP, 0.750 Recall@5\%, and six hits under the fixed validation budget. The Initial artifact decreases to 0.490 AP, 0.625 Recall@5\%, and five hits, and is therefore rejected. After validation-guided repair, AP increases to 0.518, Recall@5\% returns to 0.750, and the repaired artifact is accepted with six hits. In Fig.~\ref{fig:end-to-end-case-study}(b), this repair moves the selected segment from fold-local rank 162 to rank 124, bringing it inside the rank-128 benchmark cutoff.

The admitted segment contains a sustained capacity failure that is obscured by its high-throughput periods. Fig.~\ref{fig:end-to-end-case-study}(c) shows 109 consecutive seconds below 20 Mbps. At the highlighted prediction-overestimation event, the point prediction is 222.9 Mbps, whereas the measured mean capacity over the following 10 s is only 12.2 Mbps. The same segment produces more than 61 s of total rebuffering for every controller in Fig.~\ref{fig:end-to-end-case-study}(d) and Table~\ref{tab:end-to-end-case-study}. The case therefore connects validation feedback to an actual artifact change, a benchmark-admission decision, and measurement evidence with clear prediction-side and playback-side consequences.

\section{Conclusion}
This paper presented StarCodex, a dynamic coding harness for Starlink measurement analysis and experiment automation. We formulated the problem as a measurement-to-artifact workflow that connects measurement batches, analysis states, exposed gaps, structured coding tasks, generated artifacts, artifact acceptance, and artifact utility. StarCodex uses Codex to generate or repair analysis programs from structured coding tasks, and accepts the resulting artifacts only after code-level, data-level, measurement-level, and output-level validation. Experiments on real Starlink measurements confirm that StarCodex produces reliable analysis artifacts that improve risk discovery and carry newly exposed measurement evidence into downstream system evaluation. Overall, StarCodex establishes dynamic coding as an effective foundation for continuous network experiment automation as data and analysis requirements evolve.
Future work will extend this dynamic-coding paradigm beyond Starlink measurement analysis to broader network experimentation tasks, such as automated cross-layer diagnosis, resource-scheduling studies, and continuous benchmark maintenance under changing network conditions.

\bibliographystyle{IEEEtran}
\bibliography{ref}

\clearpage
\appendices
\setcounter{equation}{0}
\setcounter{table}{0}
\renewcommand{\theequation}{S\arabic{equation}}
\renewcommand{\thetable}{S\arabic{table}}
\input{starcodex_supplementary_content}

\end{document}

%% file: starcodex_supplementary_content.tex
\section{StarCodex Interface Examples}
\label{supp:starcodex-interface}

This appendix instantiates the StarCodex interfaces with the hard-case discovery task from outer fold~3. The running example follows one profile gap from its structured Codex instruction to an executable ranking artifact, the Harness acceptance decision, and the Experiment Products supported by the accepted artifact. The notation follows the problem formulation and framework definitions in the main text.

\subsection{Running Case for an Analysis Gap}
\label{supp:running-gap}

Consider a measurement batch $B_k$ for which the previous state $S_{k-1}$ already contains the frozen low-tail, high-volatility, high-latency, and handover-heavy profile rules. Screening replay and prediction analysis expose system-risk segments outside this coverage. StarCodex represents the missing ranking capability as a profile gap:

\smallskip
\noindent\fbox{%
\begin{minipage}{0.93\linewidth}
\scriptsize
\raggedright
\textbf{Running case: uncovered system-risk discovery}\\[0.6mm]
\textbf{Observed evidence:} some 10-min segments are replay-hard or contain prediction-overestimation evidence but are not covered by the frozen profile rules.\\
\textbf{Available inputs:} throughput, latency, and handover summary features from construction-side segments.\\
\textbf{Existing artifact:} the Registry incumbent selected by source-run-grouped validation.\\
\textbf{Missing capability:} an executable scorer that ranks future segments by uncovered system-risk likelihood.
\end{minipage}}
\smallskip

This case is represented as
\begin{equation}
\left(
B_k,\,
S_{k-1},\,
\mathcal{A}_{k-1}
\right)
\Rightarrow
g\in\mathcal{G}_k^{\mathrm{profile}} .
\label{eq:supp-running-gap}
\end{equation}
The target label is available only on the construction-side partitions. Acceptance and outer-test labels remain hidden from Codex, so the generated artifact must operate only on the allowed measurement fields.

\subsection{Structured Codex Instruction}
\label{supp:codex-instruction}

The structured coding-task fields defined in the main text are rendered into a concrete Codex instruction for this gap. The following block is a condensed version of the instruction used for outer fold~3:

\smallskip
\noindent\begin{minipage}{\linewidth}
\hrule
\vspace{1mm}
\scriptsize
\begin{verbatim}
StarCodex coding task:
  task: baseline-aware hard-case discovery
  outer_fold: 3
  gap: system-risk segments outside frozen profile rules
  safe_starting_point: Registry random-forest incumbent
  construction_data: construction.csv
  construction_target: target_uncovered_hard
  allowed_inputs:
    throughput, latency, and handover summary features
Expected artifact:
  artifact.py with build_estimator(random_state)
Required behavior:
  return an unfitted deterministic sklearn estimator;
  output one finite score per input segment.
Restrictions:
  no row, trace, source-run, region, profile-label,
  prediction-outcome, or ABR-outcome identity fields.
Harness protocol:
  one aggregate feedback report;
  commit on a separate source-run-disjoint partition.
\end{verbatim}
\vspace{-1mm}
\hrule
\end{minipage}
\smallskip

\subsection{Artifact Example and Repair Feedback}
\label{supp:artifact-lifecycle}

The generated candidate instantiates the analysis-artifact fields defined in the main text. Table~\ref{tab:supp-artifact-record} shows the concrete interface used by the Harness.

\begin{table}[!t]
\centering
\caption{Concrete hard-case discovery artifact record.}
\label{tab:supp-artifact-record}
\footnotesize
\setlength{\tabcolsep}{2.8pt}
\renewcommand{\arraystretch}{1.06}
\begin{tabular}{@{}p{0.22\linewidth}p{0.70\linewidth}@{}}
\toprule
\textbf{Artifact Field} & \textbf{Fold-3 Instantiation} \\
\midrule
Type & Hard-case discovery scorer \\
Gap $g$ & System-risk segments outside frozen profile-rule coverage \\
Input $\mathcal{I}$ & Construction table containing only the allowed measurement-summary features \\
Output $\mathcal{O}$ & One finite score for every input row, with the original row order preserved \\
Program $\pi$ & \texttt{artifact.py} exposing \texttt{build\_estimator(random\_state)} \\
Result $r$ & Ranked segment scores and the Harness acceptance record \\
\bottomrule
\end{tabular}
\end{table}

The Harness first checks that the program is executable, respects the allowed data interface, uses measurement-supported features, and returns a complete score vector. The task-specific performance gate then compares fixed-budget hits and average precision (AP) with the Registry incumbent on the source-run-disjoint acceptance partition. Fold~3 produces the following feedback and decision record:

\smallskip
\noindent\begin{minipage}{\linewidth}
\hrule
\vspace{1mm}
\scriptsize
\begin{verbatim}
Repair feedback H_{k,g}:
  retain the Registry incumbent as an anchor;
  improve fixed-budget hits first and AP second.

Acceptance gate:
  hits and AP must both be non-decreasing;
  at least one must improve over the incumbent.

Acceptance metrics (hits, recall, AP):
  Registry:  6, 0.750, 0.501
  Initial:   5, 0.625, 0.490  -> reject
  Repaired:  6, 0.750, 0.518  -> commit
\end{verbatim}
\vspace{-1mm}
\hrule
\end{minipage}
\smallskip

The repaired scorer passes the gate and is stored with its task source, program path, interface contract, and acceptance record in Artifact Memory.

\subsection{Prediction-Risk Calibration Record}
\label{supp:prediction-calibration}

The prediction-risk task uses the same acceptance interface on its source-run-disjoint calibration partition. Table~\ref{tab:supp-prediction-calibration} gives one calibration record in which the Repaired artifact is checked against the Initial artifact before the final artifact is committed.

\begin{table}[!t]
\centering
\caption{Prediction-risk artifact calibration record.}
\label{tab:supp-prediction-calibration}
\footnotesize
\setlength{\tabcolsep}{4pt}
\renewcommand{\arraystretch}{1.06}
\begin{tabular}{@{}lcc@{}}
\toprule
\textbf{Calibration Metric} & \textbf{Initial} & \textbf{Repaired} \\
\midrule
Precision@5\% & 0.2091 & 0.2067 \\
Recall@5\% & 0.7415 & 0.7331 \\
Exposure coverage@5\% & 0.5760 & 0.5625 \\
AP & 0.2897 & 0.2901 \\
\midrule
Harness decision & Commit & Reject \\
\bottomrule
\end{tabular}
\end{table}

The repaired proposal gives only a small AP change, while its fixed-budget precision, recall, and exposure coverage decrease. The paired region-stratified source-run bootstrap confidence interval for the AP difference (Repaired $-$ Initial) is $[-0.0022,\,0.0011]$, so the Harness keeps the initial artifact for the prediction-risk experiment.

\subsection{Experiment Products}
\label{supp:algorithm-output}

The accepted scorer supplies cross-fitted rankings to later experiment-compilation tasks. Those tasks produce the report, benchmark table, figure, and case-study outputs listed by the StarCodex procedure in the main text. The following examples report the final cross-fitted results.

\smallskip
\noindent\textbf{Example Experiment Report.}
\smallskip

\noindent\begin{minipage}{\linewidth}
\hrule
\vspace{1mm}
\scriptsize
\begin{verbatim}
Report: cross-fitted hard-case discovery
Accepted artifact sources:
  Initial artifact in 2 outer folds;
  Repaired artifact in 2 outer folds.
Discovery result at 5% review budget:
  49 of 56 uncovered system-risk cases;
  AP = 0.703.
Compiled benchmark:
  512 segments, 147 system-risk segments,
  55 uncovered-risk cases, 35 source runs.
\end{verbatim}
\vspace{-1mm}
\hrule
\end{minipage}
\smallskip

\noindent\textbf{Example Benchmark Table.}
\smallskip

\begin{center}
\scriptsize
\begin{tabular}{lrr}
\toprule
\textbf{Benchmark Component} & \textbf{Segments} & \textbf{Share} \\
\midrule
Uncovered system risk & 55 & 10.7\% \\
Other system risk & 92 & 18.0\% \\
Remaining selected evidence & 365 & 71.3\% \\
\midrule
Total & 512 & 100.0\% \\
\bottomrule
\end{tabular}
\end{center}

The benchmark manifest additionally stores the source run, region, measurement-profile labels, discovery score, and available prediction or replay evidence for each selected segment.

\noindent\textbf{Example Figure Product.}
\smallskip

The generated discovery figure plots uncovered-risk recall over matched review budgets and records the committed artifact source in each outer fold. The same accepted rankings are compiled into the benchmark-composition figure reported in the main text.

\noindent\textbf{Example Case Study.}
\smallskip

\noindent\begin{minipage}{\linewidth}
\hrule
\vspace{1mm}
\scriptsize
\begin{verbatim}
Case Study: artifact-repair admission in outer fold 3
Artifact transition:
  initial rank 162 -> repaired rank 124;
  benchmark cutoff = 128.
Measurement evidence:
  109 consecutive seconds below 20 Mbps.
Prediction evidence:
  point prediction = 222.9 Mbps;
  subsequent 10-s mean capacity = 12.2 Mbps.
ABR replay evidence:
  BOLA, RobustMPC, Pensieve, and SafeSABR each
  accumulate more than 61 s of rebuffering.
\end{verbatim}
\vspace{-1mm}
\hrule
\end{minipage}
\smallskip

Across the accepted artifact set, these reports, benchmark manifests, figures, and case records are concrete instances of the artifact result field $r$. Their paths and source tasks are recorded by Artifact Memory, and the resulting products update the experiment component of the analysis state.